\documentclass[twocolumn]{aastex63}

\usepackage{amsmath,amssymb,bbold,mathrsfs}
\usepackage{textcase}
\usepackage{subfigure}
\usepackage{xcolor}
\usepackage{soul}
\usepackage[nonumberlist,nosuper]{glossaries}
\setacronymstyle{long-short}

\definecolor{g}{RGB}{0,160,0}
\setstcolor{g}
\definecolor{b}{RGB}{0,0,160}
\setstcolor{b}
\definecolor{r}{RGB}{255,0,0}
\setstcolor{r}

\newcommand{\fref}[1]{Fig.~\ref{#1}}
\newcommand{\frefs}[1]{Figs.~\ref{#1}}
\newcommand{\TAUE}[1]{${\rm log_{10}}(\tau_{500})=#1$}

\newcommand{\sref}[1]{Section~\ref{#1}}

\newacronym{mo}{MO}{magneto-optical}
\newacronym{1d}{1D}{one-dimensional}
\newacronym{3d}{3D}{three-dimensional}
\newacronym{rt}{RT}{radiation transfer}
\newacronym{se}{SE}{statistical equilibrium}
\newacronym{see}{SEE}{statistical equilibrium equations}
\newacronym{clasp}{CLASP}{Chromospheric Lyman-Alpha SpectroPolarimeter}
\newacronym{tic}{HanleRT-TIC}{HanleRT Tenerife Inversion code}
\newacronym{fwhm}{FWHM}{full width at half maximum}
\newacronym{prd}{PRD}{partial frequency redistribution}
\newacronym{crd}{CRD}{complete frequency redistribution}

\received{}
\revised{}
\accepted{}

\shorttitle{Stokes-vector inversion of \ion{Mg}{2} h \& k}
\shortauthors{Li, del Pino alem\'an \& Trujillo Bueno}

\begin{document}

\title{Full Stokes-vector inversion of the solar \ion{Mg}{2} h \& k lines}

\author[0000-0001-5612-4457]{Hao Li}
\affil{Instituto de Astrof\'{\i}sica de Canarias, E-38205 La Laguna, Tenerife, Spain}
\affil{Departamento de Astrof\'\i sica, Universidad de La Laguna, E-38206 La Laguna, Tenerife, Spain}
\affil{Key Laboratory of Solar Activity and Space Weather, National Space Science Center, Chinese Academy of Sciences, Beijing, 100190, People's Republic of China}
\author[0000-0003-1465-5692]{Tanaus\'u\ del Pino Alem\'an}
\affil{Instituto de Astrof\'{\i}sica de Canarias, E-38205 La Laguna, Tenerife, Spain}
\affil{Departamento de Astrof\'\i sica, Universidad de La Laguna, E-38206 La Laguna, Tenerife, Spain}
\author[0000-0001-5131-4139]{Javier\ Trujillo Bueno}
\affil{Instituto de Astrof\'{\i}sica de Canarias, E-38205 La Laguna, Tenerife, Spain}
\affil{Departamento de Astrof\'\i sica, Universidad de La Laguna, E-38206 La Laguna, Tenerife, Spain}
\affil{Consejo Superior de Investigaciones Cient\'{\i}ficas, Spain}

\begin{abstract}
The polarization of the \ion{Mg}{2} h \& k resonance lines is the result of the 
joint action of scattering processes and the magnetic field induced Hanle, Zeeman, 
and magneto-optical effects, thus holding significant potential for the diagnostic 
of the magnetic field in the solar chromosphere. The Chromospheric LAyer 
Spectro-Polarimeter sounding rocket experiment, carried out in 2019, successfully 
measured at each position along the 196 arcsec spectrograph slit the wavelength 
variation of the four Stokes parameters in the spectral region of this doublet 
around $280$~nm, both in an active region plage and in a quiet region close to the 
limb. We consider some of these CLASP2 Stokes profiles and apply to them the 
recently-developed HanleRT Tenerife Inversion Code, which assumes a one-dimensional 
model atmosphere for each spatial pixel under consideration (i.e., it neglects 
the effects of horizontal radiative transfer). We find that the non-magnetic 
causes of symmetry breaking, due to the horizontal inhomogeneities and the 
gradients of the horizontal components of the macroscopic velocity in the solar 
atmosphere, have a significant impact on the linear polarization profiles. By 
introducing such non-magnetic causes of symmetry breaking as parameters in our 
inversion code, we can successfully fit the Stokes profiles and provide an 
estimation of the magnetic field vector. For example, in the quiet region pixels, 
where no circular polarization signal is detected, we find that the magnetic field 
strength in the upper chromosphere varies between 1 and 20 gauss.  
\end{abstract}


\section{Introduction} \label{intro}

The magnetic field permeating the solar atmosphere plays a key role in its structure,
energy transfer, and the eventual eruptive phenomena. The inference of the magnetic
field in the photosphere usually relies on the use of inversion codes of polarization
signals magnetically induced by the Zeeman effect
\citep{Iniesta2016LRSP,Lagg2017SSRv,delaCruz2017SSRv}.
The amplitude of the circular polarization caused by the Zeeman effect scales with the
ratio (generally smaller than unity) between the Zeeman splitting and the Doppler 
line width, while the linear polarization signals scale with the square of this 
quantity \citep{LL04}. Chromospheric lines tend to be wider than photospheric lines, 
and the magnetic field generally decreases with height. For these reasons, the inference 
of chromospheric magnetic fields via the Zeeman effect becomes 
considerably more challenging, except in active regions where its circular 
polarization signals can be measured even in strong ultraviolet lines like 
\ion{Mg}{2} h \& k \citep{Ishikawa2021SA,Li2023ApJ}.

The Hanle effect is the magnetically induced modification of the linear polarization 
caused by the scattering of anisotropic radiation in a spectral line and holds 
significant potential for the diagnostic of magnetic fields, albeit being 
significantly more difficult to exploit \citep[e.g., the monograph by][]{LL04}. 
Motivated by the theoretical investigations on the polarization caused by 
scattering processes and the Hanle and Zeeman effects in the \ion{Mg}{2} h \& k 
lines around 280~nm (\citealt{Belluzzi2012ApJa,AlsinaBallester2016ApJ,Tanausu2016ApJ}, 
and \citealt{Tanausu2020ApJ}) the Chromospheric LAyer SpectroPolarimeter (CLASP2;
\citealt{Narukage2016SPIE,Song2018SPIE}) suborbital rocket experiment was carried 
in 2019 with the aim of observing the four Stokes parameters in this ultraviolet 
spectral region and inferring the magnetic field vector in the chromosphere. 
The mission was successful and obtained unprecedented spectropolarimetric data 
of this spectral region in both a quiet region close to the solar limb 
\citep{Rachmeler2022ApJ} and in an active region plage \citep{Ishikawa2021SA}. 
The CLASP2 spectropolarimetric observations confirmed the theoretical predictions 
based on the quantum theory of spectral line 
polarization \citep[see the review by][]{TrujilloBueno2022ARA&A}, 
which showed that the combined action of \gls*{prd} effects and quantum mechanical
interference between the magnetic sublevels pertaining to the two upper $J$-levels
of the \ion{Mg}{2} h \& k lines (hereafter, $J$-state interference) produce sizable
scattering polarization signals in the near and far wings of these lines
\citep{Belluzzi2012ApJa}, and that such signals are sensitive to the magnetic field
via the \gls*{mo} terms of the Stokes-vector transfer equation
\citep{AlsinaBallester2016ApJ,Tanausu2016ApJ}.

While there are well-known Stokes inversion codes like HAZEL 
\citep{AsensioRamos2008ApJ} which exploit scattering polarization and the Hanle 
and Zeeman effects assuming \gls*{crd} and a constant-property slab of plasma 
levitating at a given height above the solar visible disk, until very recently 
there was no inversion code capable of tackling the inversion of Stokes profiles 
caused by the joint action of all the above-mentioned effects in strong 
chromospheric lines, like those of the \ion{Mg}{2} h \& k doublet. These 
resonance lines show both significant \gls*{prd} and \gls*{rt} effects, and 
their forward modeling requires solving the problem of the generation and transfer 
of polarized radiation out of thermodynamic equilibrium (non-LTE) accounting
for \gls*{prd}, scattering polarization, $J$-state interference, and the Hanle, 
Zeeman, and \gls*{mo} effects in the general magnetic field regime (incomplete 
Paschen-Back regime) in an optically thick plasma. In order to tackle this 
inversion problem, we recently developed the Stokes \gls*{tic} \citep{Li2022ApJ}, 
based on the HanleRT forward solver 
\citep{Tanausu2016ApJ,Tanausu2020ApJ}.\footnote{HanleRT-TIC is publicly available
at \url{https://gitlab.com/TdPA/hanlert-tic}.}

The intensity and circular polarization profiles observed by CLASP2 in an active 
region plage have been exploited by applying the weak field approximation
\citep{Ishikawa2021SA,AfonsoDelgado2023ApJ} and the \gls*{tic} inversion code
\citep{Li2023ApJ}, obtaining an unprecedented map of the longitudinal component 
of the magnetic field from the photosphere to the upper chromosphere, just below 
the transition region. The linear polarization observed by CLASP2 across the 
\ion{Mg}{2} h \& k lines was shown in \cite{Rachmeler2022ApJ}, comparing the 
CLASP2 observations of the quiet Sun target with the theoretical predictions of 
\cite{Belluzzi2012ApJa} in an unmagnetized semi-empirical model of the quiet 
solar atmosphere; the authors argued that horizontal inhomogeneities
and magnetic fields are needed to explain the observed signals and their spatial
variations.

In a \gls*{1d} model atmosphere, static or without horizontal components 
of the plasma macroscopic velocity, the only way to break the axial symmetry 
of the radiation field that illuminates each spatial point within the medium is 
by the presence of a magnetic field inclined with respect to the vertical axis, 
the axis along which the model's physical quantities vary. However, at each height 
in the real solar atmosphere we have horizontal inhomogeneities in the plasma 
temperature and density, as well as macroscopic motions with spatial gradients also 
in the horizontal component of the velocity. Therefore, in general, in the real 
solar atmosphere the radiation field does not have axial symmetry around the solar 
radius vector through the spatial point under consideration, even in the absence 
of a magnetic field \citep[e.g.,][]{Jaumeetal2021}. The impact of such non-magnetic 
causes of symmetry breaking on the linear polarization caused by the scattering of 
anisotropic radiation in the \ion{Mg}{2} h \& k lines can be modeled applying a 
three-dimensional (3D) RT code, but the development of such a code with PRD and 
$J$-state interference capable of performing calculations in realistic 3D models 
with today's supercomputer facilities is still an unachieved 
challenge.\footnote{3D radiative transfer codes for the synthesis
\citep{Stepan2013A&A} and inversion \citep{Stepan2022A&A,Stepanetal2024} of
Stokes profiles accounting for atomic polarization exist, but they use the
CRD approximation, which is not suitable for modeling the \ion{Mg}{2} h \& k
lines.} Our \gls*{tic} takes into account the effects of PRD and $J$-state
interference in the presence of arbitrary magnetic fields, but ignoring the effects
of horizontal radiative transfer (i.e., \gls*{tic} is a 1D plane-parallel RT code
for the synthesis and inversion of Stokes profiles). The only way of bypassing this
issue in 1D geometry, i.e., without solving the full 3D RT problem, is by parameterizing
this missing contribution to the lack of axial symmetry. \gls*{tic} allows such a
functionality.

Our first step in this work is to estimate the magnetic field vector in the 
chromosphere through the application of the \gls*{tic} to some of the Stokes 
profiles of the \ion{Mg}{2} h \& k lines observed by CLASP2. To this end, after 
briefly describing in \sref{sec2} the CLASP2 observations, in \sref{sec3}
we describe our parameterized approach to account for the impact of the above-mentioned
non-magnetic causes of symmetry breaking in our pixel by pixel inversions 
with \gls*{tic}. The inversion of four representative \ion{Mg}{2} h \& k Stokes 
profiles are shown in \sref{sec4}. \sref{sec5} discusses possible degeneracies
introduced by our parameterization of the non-magnetic causes of symmetry breaking 
and how they affect the inferred magnetic field. Finally, \sref{sec6} summarizes 
our main conclusions.


\section{Summary of the CLASP2 observations} \label{sec2}


\begin{figure*}[htp]
  \center
  \includegraphics[width=1.0\textwidth]{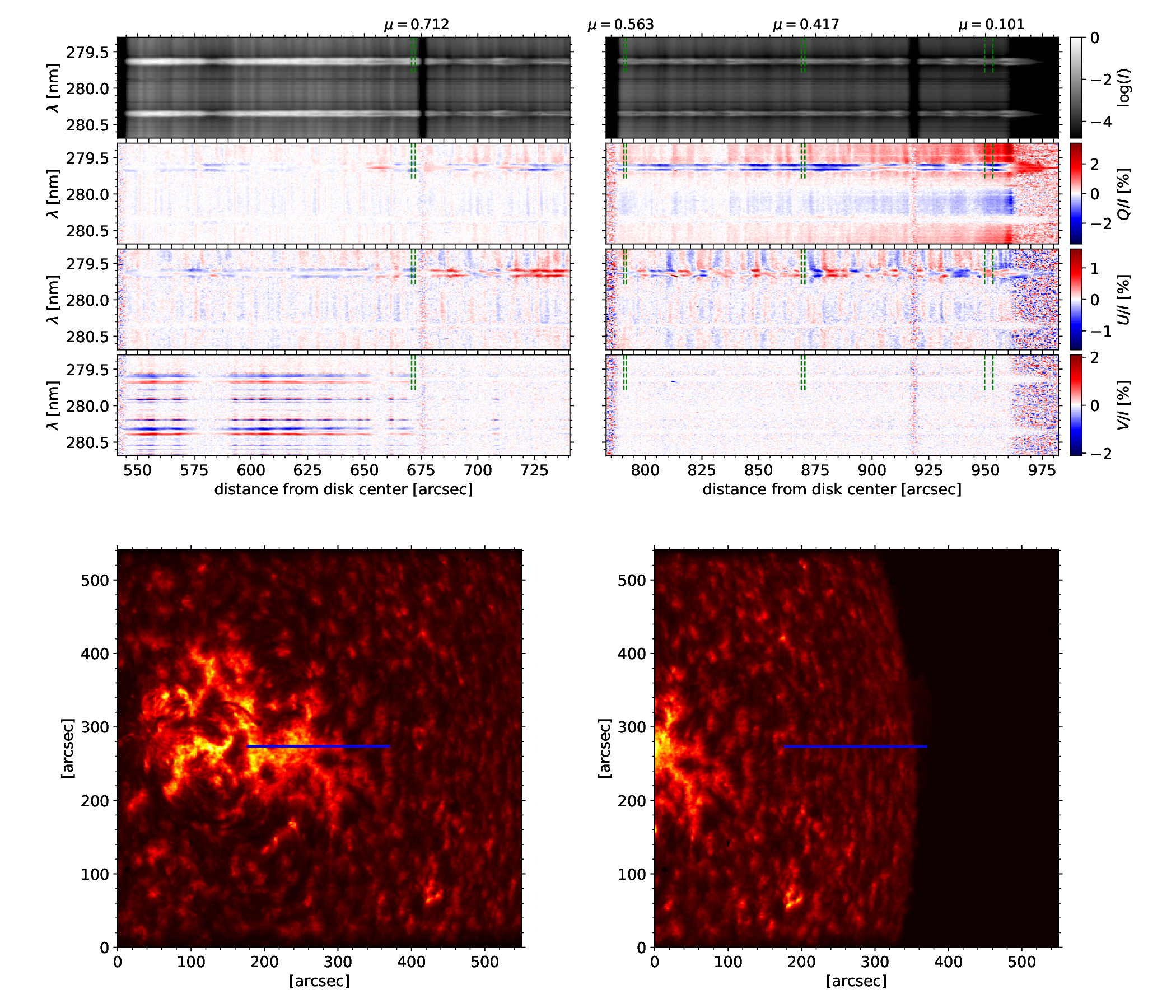}
  \caption{In the top panels, intensity (in log-scale and normalized to the maximum) and Stokes 
  fractional polarization ($Q/I$, $U/I$, and $V/I$) profiles obtained by the CLASP2.
  In the bottom panels, slit-jaw images in the \ion{H}{1} Lyman-$\alpha$ line recorded
  by CLASP2.
  The left and right columns correspond to the observations of the 
  plage and limb targets, respectively. The regions delimited by green dashed 
  lines in the top panels indicate the locations of the 4 spatially averaged profiles in 
  \frefs{fig2}--\ref{fig5}.
  The blue lines in the bottom panels indicate the position of the spectrograph's slit.
 }
  \label{fig1}
\end{figure*}

The data used in this work were obtained by the CLASP2 suborbital space mission 
on 11 April 2019. Sit-and-stare observations were carried out with the 196~arcsec 
spectrograph slit located at three consecutive positions on the solar disk, namely 
the disk center, an active region plage at the east side of NOAA~12738 with a 
nearby enhanced network, and a quiet Sun region near the north-east limb. The 
plage and quiet-Sun target observations were obtained between 16:53:40 and 
16:56:16~UT (156~s) and between 16:56:25 and 16:58:45~UT (140~s), respectively 
\citep{Ishikawa2021SA}. The temporally averaged signals have a polarization 
accuracy better than 0.1\% \citep{Song2022SoPh}. 

The spectral range of CLASP2 is between 279.30 and 280.68~nm, with a sampling of
49.9~m\AA/pixel. The spectral point spread function can be approximated with a 
gaussian with a full width at half maximum (FWHM) of 110~m\AA\ 
\citep{Song2018SPIE,Tsuzuki2020SPIE}. The CLASP2 spectral range contains 
the \ion{Mg}{2} h and k lines at 279.6 and 280.3~nm, respectively, two 
\ion{Mg}{2} blended transitions at 279.88~nm, whose lower level are the 
upper levels of the h \& k lines, as well as several other lines (e.g., the 
\ion{Mn}{1} resonance lines at 279.91 and 280.19 nm). 

The temporally averaged observations of the plage and quiet-Sun limb targets are 
shown in \fref{fig1}. The circular polarization caused by the Zeeman effect is 
clearly prominent in about 2/3 of the plage target (between 540 and 670 arcsec), 
faint in the nearby enhanced network (between 670 and 740 arcsec), and non-detected 
in the quiet region closer to the limb (right panels in \fref{fig1}). The linear 
polarization follows almost the opposite behavior, with weaker scattering 
polarization signals within the plage region due to the Hanle and MO depolarization 
caused by its stronger magnetic fields. Although in the quiet region the circular 
polarization was not detected, the estimation of the magnetic field vector is 
still possible via the Hanle effect in the Mg {\sc ii} k line. Given the 
significant computational demand of full Stokes inversions with scattering 
polarization and the novelty of the inversion approach in this work, here we 
focus on four representative profiles located at heliocentric angles, $\theta$, with 
$\mu = \cos{\theta} = 0.712$, 0.563, 0.417, and 0.101.
At each of these locations, we have spatially averaged the observed Stokes
profiles at a few adjacent pixels (3, 2, 3, and 7 pixels, respectively)
sharing similar profiles, increasing the signal to noise ratio without causing
significant cancellations of the observed polarization signals. We
show as example the Stokes profiles for the individual pixels included in the
average at $\mu=0.712$ in Appendix~\ref{app-avg}.


\section{The Stokes inversion strategy} \label{sec3}

In this section we describe the inversion strategy we have applied to infer 
the magnetic field vector from the observed Stokes profiles, showing illustrative 
results for a point on the plage target and at three locations on the quiet Sun 
target. In particular, we explain the parameterization we have included in the 
\gls*{tic} in order to account for the possible impact of the non-magnetic 
causes of symmetry breaking due to the presence of horizontal inhomogeneities 
and macroscopic horizontal motions.

\subsection{Parameterization of the axial symmetry breaking} \label{sec3.2}

At each iterative step needed for solving the non-LTE Stokes inversion problem 
the \gls*{tic} solves the non-LTE spectral synthesis problem of the generation 
and transfer of polarized radiation assuming \gls*{1d} plane-parallel geometry. 
Therefore, at each pixel the inversion of the emergent Stokes profiles is done 
assuming that there is no radiative interaction with the surrounding pixels. 
In other words, the effects of horizontal RT in the solar atmosphere (because 
it is a 3D plasma) are neglected. With such an approximation in \gls*{tic} any 
breaking of the axial symmetry of the pumping radiation field would be due to 
a non-vertical magnetic field vector and/or to vertical gradients in the 
horizontal components of the macroscopic velocity, because at each height in a 
\gls*{1d} plane-parallel model all the quantities are assumed to be the same
along the horizontal directions. However, 
in the solar atmosphere, or in a \gls*{3d} model atmosphere, the horizontal RT 
effects resulting from the horizontal inhomogeneities in the physical properties 
of the plasma break the axial symmetry of the radiation field that illuminates 
each spatial point within the medium. Such non-magnetic causes of symmetry 
breaking are important, because they can have a significant impact on the 
linear polarization caused by the scattering of anisotropic radiation in a 
spectral line \citep[e.g.,][]{MansoTrujillo2011,delPinoetal2018,Jaumeetal2021}.

The degree of axial symmetry breaking in the radiation field illuminating each 
spatial point within a model atmosphere is quantified by the $Q\neq0$ components 
of the $J^K_Q$ radiation field tensor \citep[for a detailed derivation and 
description of these tensors, see][]{LL04}. As explained above, when solving 
the spectral synthesis problem pixel by pixel assuming \gls*{1d} plane-parallel 
geometry we miss contributions to these components of the radiation field tensor 
that can be important for the linear polarization. What we propose in this work 
is to parameterize the missing contributions as ad-hoc contributions to the 
radiation field tensors in \gls*{tic}, by defining a new set of $J'^K_Q$ tensor 
components as

\begin{subequations}
\begin{align}
J'^2_1 & = J^2_1 + {J^\dagger}^2_1 , \label{Eq0a} \\
J'^2_2 & = J^2_2 + {J^\dagger}^2_2 , \label{Eq0b}
\end{align}
\label{Eq0}
\end{subequations}
where $J^2_1$ and $J^2_2$ are the tensor components calculated by integrating 
the Stokes parameters \citep[see][]{LL04} resulting from the solution of the 
\gls*{rt} equations, $J'^2_1$ and $J'^2_2$ are the final values that go into 
the \gls*{see} and the emissivity in the \gls*{rt} equations, and 
${J^\dagger}^2_1$ and ${J^\dagger}^2_2$ are the ad-hoc parameterized additional 
contributions. We do not consider ad-hoc contributions to the $J^1_1$ component 
because it does not impact the emergent linear polarization profiles. The ad-hoc 
components of the radiation field tensor are then

\begin{subequations}
\begin{align}
{J^\dagger}^2_1 & = r_1J^0_0 + i_1J^0_0 i \label{Eq1a} , \\
{J^\dagger}^2_2 & = r_2J^0_0 + i_2J^0_0 i,\label{Eq1b} ,
\end{align}
\label{Eq1}
\end{subequations}
where $i$ is the imaginary unit and $J^0_0$ is the component of the radiation 
field tensor corresponding to the mean intensity, calculated by integrating the 
Stokes $I$ parameter resulting from the solution of the \gls*{rt} equations 
\citep[see][]{LL04}. The coefficients $r_1$, $i_1$, $r_2$, and $i_2$ are
parameters in the inversion. The relation between these quantities and the Stokes 
parameters for the linear polarization has been previously exploited by 
\citet{Zeuner2020ApJ,Zeuneretal2024} in their observational study of the 
scattering polarization in the \ion{Sr}{1} photospheric line at 4607~\AA. 
Hereafter, we use $J^\dagger$ to refer to the set of four ad-hoc contributions 
of the radiation field tensor, $r_1$, $i_1$, $r_2$, and $i_2$, which we quantify 
below in percentage of the $J^0_0$ component.


\subsection{Inversion cycles and nodes} \label{sec3.3}

We have applied an inversion strategy similar to that outlined in 
\citet{Li2023ApJ}, with two cycles to invert the temperature ($T$), the bulk 
vertical velocity ($v_z$), the micro-turbulent velocity ($v_{\rm turb}$), and 
the gas pressure at the top boundary ($P_{\rm g}$) from the Stokes $I$ profile 
(hereafter, the non-magnetic cycles). The first cycle has four nodes in $T$ and 
three in $v_z$ and $v_{\rm turb}$, while the second one has seven and four, 
respectively. In these non-magnetic cycles the velocity is assumed to be 
parallel to the local vertical because it significantly decreases the
computational demands. Once the Stokes $I$ profiles are fitted, $T$, $v_z$,
$v_{\rm turb}$, and $P_{\rm g}$ are fixed.

In the subsequent cycles we retrieve the magnetic field and $J^\dagger$ from 
Stokes $Q$, $U$, and $V$ (hereafter, the magnetic cycles). The Stokes inversion 
gives us the longitudinal component of the magnetic field ($B_\parallel$), the 
transverse component of the magnetic field with respect to the line of sight (LOS; 
$B_\perp$), and the azimuth of the magnetic field vector in the plane 
perpendicular to the LOS ($\phi_{B_\perp}$). We use these components 
of the magnetic field vector because Stokes $U$ in the wings of the \ion{Mg}{2} 
h \& k lines is sensitive to the sign of $B_\parallel$ through the \gls*{mo} 
effects, although the addition of $J^\dagger$ can alter this dependency. We
also provide the magnetic field strength ($B$), the magnetic field inclination
($\theta_B$) with respect to the local vertical, and the azimuth of the magnetic
field ($\chi_B$) in the local reference frame because it can be helpful to study 
the Hanle effect ambiguities. The details of the number of cycles and nodes are 
different for each of the four selected Stokes profiles, and they are described 
in Sec.~\ref{sec4}.

In all inversion cycles the spectral syntheses are performed in a model 
atmosphere with 60 non-equally spaced layers between \TAUE{-8.0} and $1.0$. 
The errors are computed from the diagonal of the Hessian matrix 
\citep[see][]{Li2022ApJ}.


\section{Results}\label{sec4}

\begin{figure*}[htp]
  \center
  \includegraphics[width=1.0\textwidth]{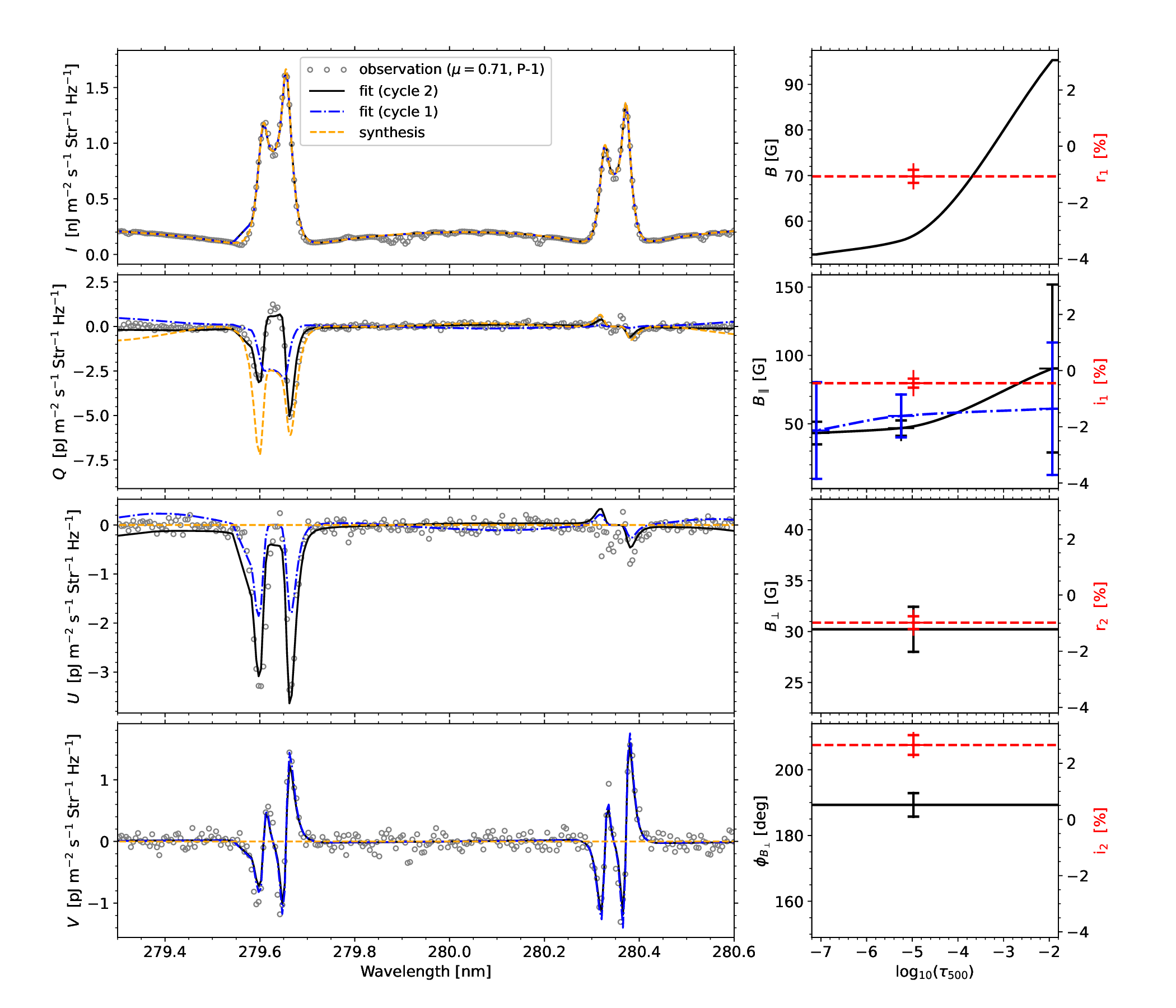}
  \caption{\textbf{Left column}: from top to bottom, Stokes $I$, $Q$, $U$, and 
  $V$ profiles, respectively. The open circles correspond to the temporally and 
  spatially averaged observed Stokes profiles. The dash-dotted blue and solid 
  black curves are the inversion fit from the first and the second magnetic cycles, 
  respectively, while the dashed orange curves show the profile synthesized in the 
  inverted atmosphere neglecting the magnetic field. 
  Note that the $k_{\rm 1v}$ minimum at around 279.57~nm is blended with
  a \ion{Mn}{1} resonance line and that the corresponding wavelength points are
   not considered in the inversion, resulting in a straight line in the fit.
  \textbf{Right column}: from 
  top to bottom, magnetic field strength, magnetic field longitudinal component, 
  magnetic field transversal component, and magnetic field azimuth in the plane 
  perpendicular to the LOS, in solid black (dash-dotted blue) curves from
  the second (first) magnetic cycle (left axis), respectively, and the real part 
  of the $Q=1$ tensor, the imaginary part of the $Q=1$ tensor, the real part of 
  the $Q=2$ tensor, and the imaginary part of the $Q=2$ tensor characterizing the 
  lack of axial symmetry, in dashed red curves (right axis), respectively. This 
  case corresponds to P-1, a region at the edge of the plage
  ($\mu=0.712$ in \fref{fig1}).}
  \label{fig2}
\end{figure*}
  
\begin{figure*}[htp]
  \center
  \includegraphics[width=1.0\textwidth]{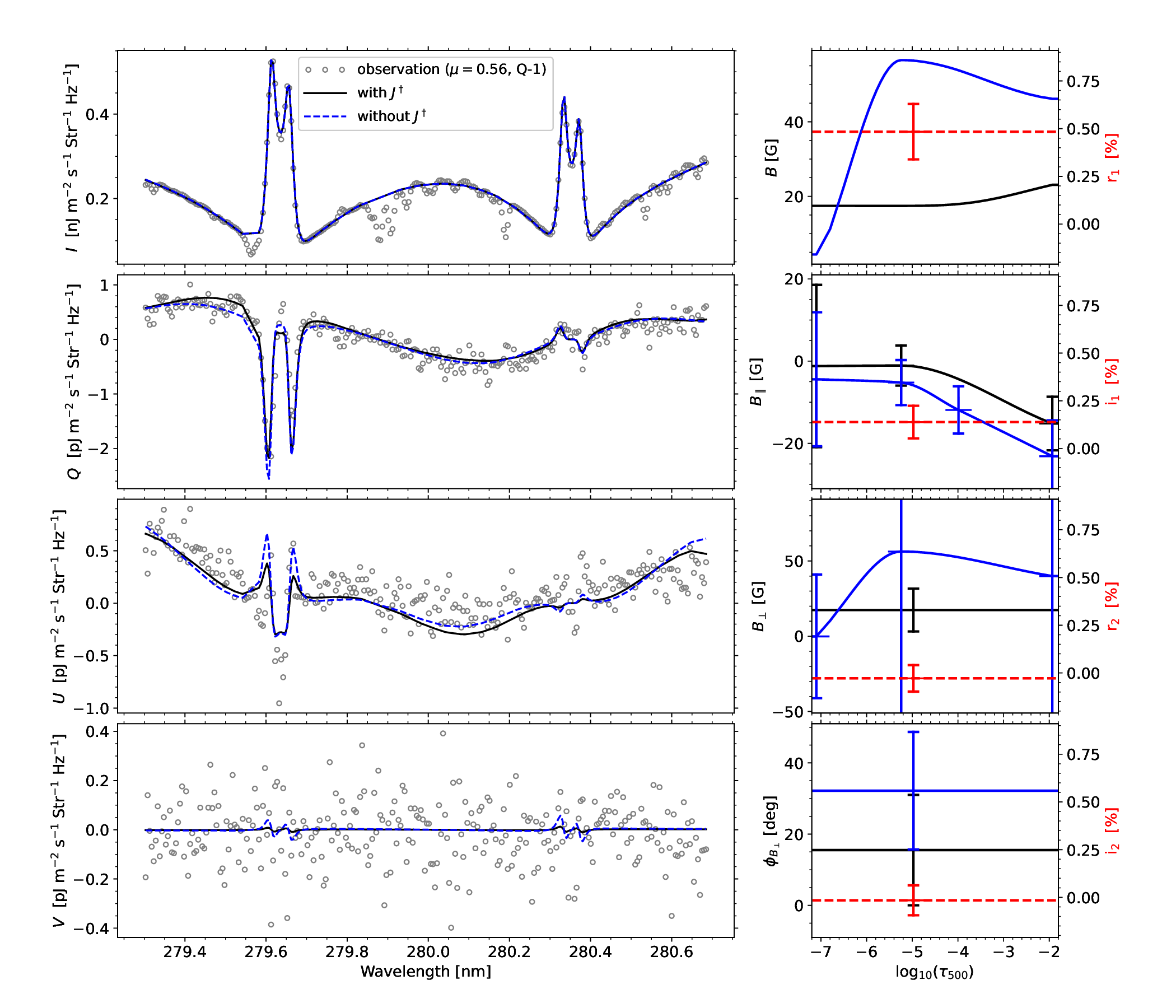}
  \caption{\textbf{Left column}: from top to bottom, Stokes $I$, $Q$, $U$, and 
  $V$ profiles, respectively. The open circles correspond to the temporally and 
  spatially averaged observed Stokes profiles. The solid black (blue) curves are 
  the inversion fit when accounting for (neglecting) the parameters for the 
  horizontal inhomogeneity. \textbf{Right column}: from top to bottom, magnetic 
  field strength, magnetic field longitudinal component, magnetic field 
  transversal component, and magnetic field azimuth in the plane perpendicular 
  to the LOS, in solid black (blue) curves for the inversion accounting 
  for (neglecting) the parameters for the horizontal inhomogeneity (left axis), 
  respectively, and the real part of the $Q=1$ tensor, the imaginary part of 
  the $Q=1$ tensor, the real part of the $Q=2$ tensor, and the imaginary part 
  of the $Q=2$ tensor characterizing the lack of axial symmetry, in dashed red 
  curves (right axis), respectively. This case corresponds to Q-1,
   a region in the quiet Sun ($\mu=0.563$ in \fref{fig1}).}
  \label{fig3}
\end{figure*}
  
\begin{figure*}[htp]
  \center
  \includegraphics[width=1.0\textwidth]{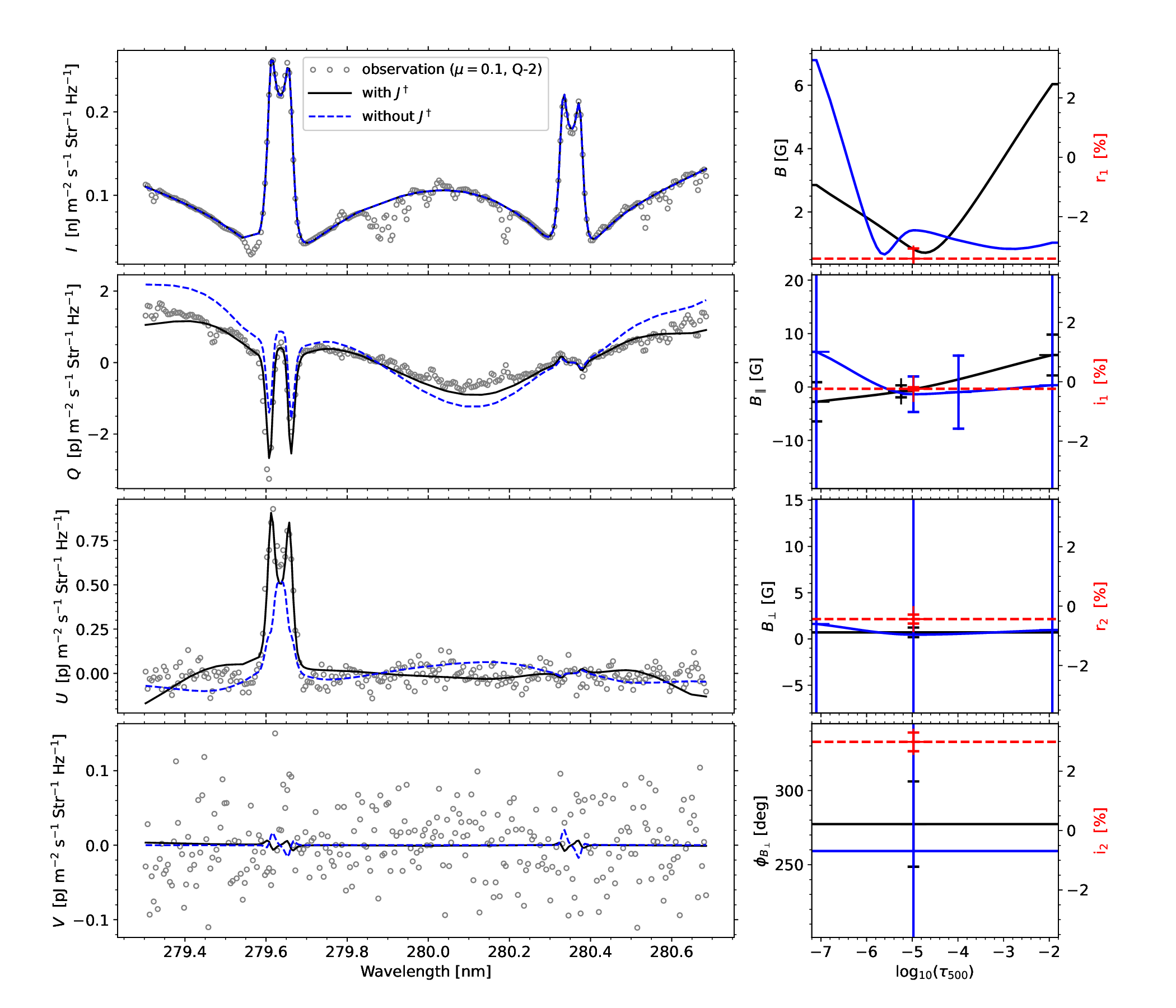}
  \caption{Same as \fref{fig3}, but for Q-2, the region in the quiet
  Sun with $\mu=0.101$ (see \fref{fig1}).}
  \label{fig4}
\end{figure*}
  
\begin{figure*}[htp]
  \center
  \includegraphics[width=1.0\textwidth]{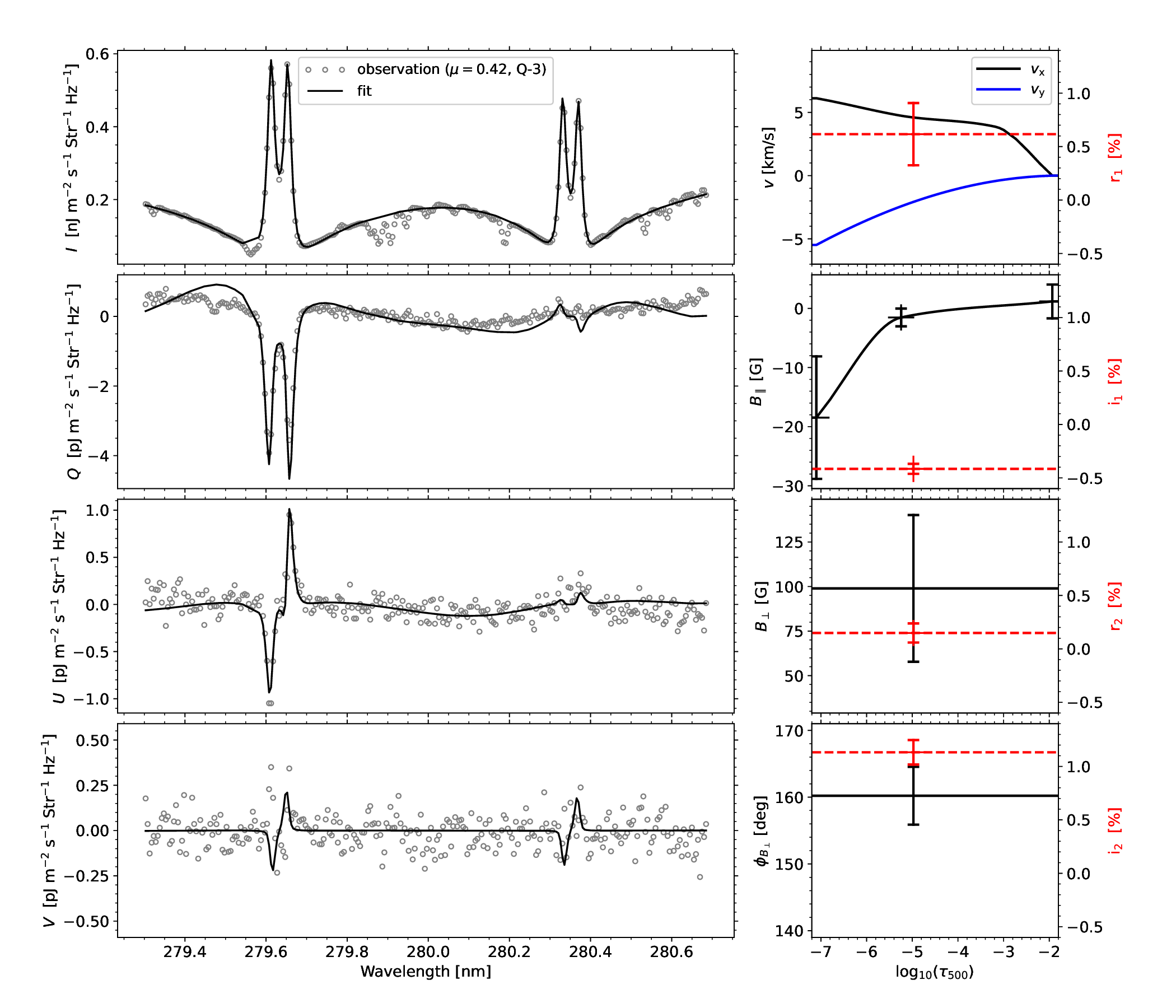}
  \caption{\textbf{Left column}: from top to bottom, Stokes $I$, $Q$, $U$, and 
  $V$ profiles, respectively. The open circles correspond to the temporally and 
  spatially averaged observed Stokes profiles. The solid black curves are the 
  inversion fit. \textbf{Right column}: from top to bottom, horizontal component 
  of the plasma velocity, magnetic field longitudinal component, magnetic field 
  transversal component, and magnetic field azimuth in the plane perpendicular 
  to the LOS, in solid black curves (left axis), respectively, and the 
  real part of the $Q=1$ tensor, the imaginary part of the $Q=1$ tensor, the 
  real part of the $Q=2$ tensor, and the imaginary part of the $Q=2$ tensor 
  characterizing the lack of axial symmetry, in dashed red curves (right axis), 
  respectively. The black solid curve in the velocity panel corresponds to the 
  component in the plane containing the LOS and the local vertical 
  (x component), while the blue solid curve corresponds to the component 
  perpendicular to the same plane (y component). This case corresponds to Q-3, a 
  region in the quiet Sun ($\mu=0.417$ in \fref{fig1}).}
  \label{fig5}
\end{figure*}

In this section we show the results of applying the \gls*{tic} to invert the 
Stokes profiles observed by CLASP2 at four representative locations, one just 
at the edge of the plage target (where the circular polarization caused by the 
Zeeman effect was detected, hereafter P-1) and three positions in the quiet-Sun target (where 
the measured circular polarization was at the noise level). In all cases, the 
Stokes $Q$ and $U$ profiles were detected, which result from the combined action 
of scattering processes and the Hanle and MO effects. As shown below, in one of 
the chosen locations the Stokes profiles could be fitted without introducing 
$J^\dagger$ (hereafter Q-1), while in another location the Stokes 
profiles could only be fitted 
with $J^\dagger$ (i.e., by acknowledging a non-magnetic cause of axial symmetry 
breaking, hereafter Q-2). Interestingly, the Stokes profiles of one of the 
chosen locations could only be fitted via the axial symmetry breaking that 
results from a vertical gradient in the horizontal component of the 
macroscopic velocity (hereafter Q-3).


\subsection{Stokes profiles at the edge of the plage (P-1)} \label{sec4.1}

The open circles of \fref{fig2} show the CLASP2 Stokes profiles at the edge of 
the plage target (see the $\mu=0.712$ location in \fref{fig1}), where we have 
significant linear and circular polarization signals. The orange dashed curves 
in \fref{fig2} show the Stokes profiles calculated in the model atmosphere that 
results from the first two cycles of the inversion (i.e., without magnetic field 
or $J^\dagger$). As expected, due to the axial symmetry of the model and the 
absence of magnetic field, Stokes $U$ and $V$ are zero. Notably, the right 
trough of the k line in the synthetic Stokes $Q$ profile shows a similar 
amplitude to the observation. Typically, the impact of the Hanle and/or MO 
effects is to depolarize and rotate the linear polarization ($Q\rightarrow U$ 
in this case). It is thus impossible to find a magnetic field vector such that 
all Stokes $Q$, $U$, and $V$ profiles are fitted at the same time. The addition 
of the $J^\dagger$ contribution is necessary to achieve a good fit. This is not 
surprising, because at the edge of the plage we can expect a significant lack of
axial symmetry, as seen in Fig.~2 of \cite{Ishikawa2023ApJ}.

We invert the polarization profiles in two magnetic cycles. In the first cycle 
we retrieve the longitudinal component of the magnetic field with 3 nodes
from only Stokes $V$. In this step we assumme that the magnetic field is vertical
so the problem is still axially symmetric, allowing for 
the determination of a better initial guess of the magnetic field without too 
heavy computational requirements. In the second cycle we infer $B_\parallel$ 
with three nodes, and $B_\perp$, $\phi_{B_\perp}$, $r_1$, $i_1$, $r_2$, and 
$i_2$ with one node. Although a constant value for these quantities is not 
realistic, if we manage to fit the observed Stokes profiles it means that there 
is not enough information to recover their gradient, and it is preferable to keep 
the compatible model with the minimum degrees of freedom. Moreover, a small 
number of degrees of freedom results in a lower computational demand
and, generally, it helps to avoid local minima in the inversion procedure.

The black curves in the left column of \fref{fig2} show the best fit from the 
full inversion. The magnetic field strength and the LOS components of 
the magnetic field vector, as well as the value of $J^\dagger$, are shown in 
the right column of the figure. The magnetic field strength decreases from 
about 60~G at \TAUE{-4.0} to about 55~G at \TAUE{-7.0}. The $J^\dagger$ values 
lie between half of a percent and a few percent. The blue curves show the 
inversion results after the first magnetic cycle, which fits  only Stokes $V$. 
Although the inferred $B_\parallel$ values are somewhat different from those 
from the full Stokes inversion, the values are compatible within the error bars 
and the Stokes $V$ fits are of similar quality. Since $B_\parallel$ also 
impacts the linear polarization via the \gls*{mo} effect, the full Stokes 
inversion can better constrain $B_\parallel$. 


\subsection{Quiet Sun Stokes profiles (Q-1)} 

Figure~\ref{fig3} shows the Stokes profiles of the \ion{Mg}{2} h \& k lines 
observed by CLASP2 in the quiet Sun target at $\mu = 0.563$ 
(see \fref{fig1}). The circular polarization caused by the Zeeman effect is 
at the noise level, but the linear polarization due to scattering processes 
is significant, especially in the line wings.

We applied two magnetic cycles to invert these Stokes profiles. In contrast 
with the plage target profiles studied in Sec.~\ref{sec4.1} (P-1), the circular 
polarization is now at the noise level, reason why we cannot obtain a first 
estimation of $B_\parallel$ from Stokes $V$. Instead, in the first magnetic 
cycle we invert the Stokes $Q$, $U$, and $V$ parameters to infer $B_\parallel$ 
with three nodes and $B_\perp$ and $\phi_{B_\perp}$ with one node. 
We point out that the linear polarization in the wings of the
\ion{Mg}{2} h \& k lines is sensitive to $B_\parallel$, and it is thus
necessary to sucessfully fit the observation. Including the Stokes $V$
profile in the inversion, even if the observation shows that the signal is
below the noise level, adds a good constraint to the upper value of $B_\parallel$.
For the initialization we used $B_\parallel=0$, $B_\perp=10$~G and 
$\phi_{B_\perp}=2.5$~rad (i.e., about $143^\circ$). The initial value of 
$B_\perp$ is chosen because the Hanle effect at the center of the k line 
is sensitive to magnetic fields with strengths between approximately 5 and 
100~G, so we initialize the magnetic field in the region of sensitivity. 
Afterwards, we perform inversions with two extra different cycles. The blue 
dashed curves in \fref{fig3} show the result of an inversion with four nodes 
for $B_\parallel$, and three nodes for $B_\perp$ and $\phi_{B_\perp}$.
The black solid curves in \fref{fig3} show the result of an inversion keeping 
the same number of nodes in $B_\parallel$, $B_\perp$, and $\phi_{B_\perp}$, 
but adding one node for $r_1$, $i_1$, $r_2$, and $i_2$.

As can be seen by comparing the black and blue curves of \fref{fig3}, the 
CLASP2 Stokes profiles at the $\mu = 0.563$ quiet-Sun target location can be 
successfully fitted with or without the $J^\dagger$ contribution. Although 
there are evident differences at the plot level, the values of the cost 
function are similar in both inversions. When the $J^\dagger$ contribution 
is not used in the Stokes inversion, the inferred model shows a steep 
stratification in $B_\perp$ with quite large errors. The inversion with 
$J^\dagger$ gives instead a smoother stratification of $B_\perp$ with smaller 
error bars. The $J^\dagger$ values are less than one percent, and the $r_2$ 
and $i_2$ components are very close to zero, indicating a relatively weak 
potential contribution from \gls*{3d} effects. In any case, both solutions 
are compatible within the error bars for the noise of the CLASP2 observation, 
so the differences between the two Stokes inversions we have just described 
is likely representative of the degree of uncertainty.

\subsection{Quiet Sun Stokes profiles (Q-2)} 

Figure~\ref{fig4} shows the Stokes profiles observed by CLASP2 in the quiet 
Sun target at $\mu=0.10$. We have performed two inversions with the same nodes 
and cycles as described above for the Stokes profiles in \fref{fig3}. In this 
case, neglecting the contribution of $J^\dagger$ cannot provide a good fit to 
the observed Stokes profiles. The wings of Stokes $Q$ and $U$ are sensitive to 
the presence of magnetic fields via the MO effects. Because the observed Stokes 
$U$ at this quiet-Sun location is negligible in the far wings, the inversion 
without $J^\dagger$ predicts $B_\parallel=0$ in the upper photosphere, where
the far wings originate. However, this constraint is incompatible with the 
far wings of the observed Stokes $Q$, which cannot be fitted simultaneously 
(it would be necessary to add $B_\parallel$, which in turn would produce a 
miss-fit in the Stokes $U$ far wings). In addition, this inversion is not able 
to even reproduce the shape of the \ion{Mg}{2} k line in Stokes $U$. It is thus 
clear that it is necessary to include $J^\dagger$ in order to be able to fit 
the Stokes profiles observed at this quiet-Sun location.

The $J^\dagger$ values shown in \fref{fig4}, of several percent, indicate a 
significant symmetry breaking contribution from \gls*{3d} effects. In this 
inversion the inferred magnetic field is rather weak, of the order of few 
gauss. Even relatively weak magnetic fields can leave observable imprints 
in the linear polarization of the \ion{Mg}{2} h \& k lines. Note, however, 
that the lower limit to this sensitivity is around 2 gauss, which is 
approximately one tenth of the critical Hanle field for the \ion{Mg}{2} k line.


\subsection{Quiet Sun Stokes profiles (Q-3)} 

At some slit locations in the quiet Sun target, the CLASP2 observations showed 
antisymmetric Stokes $U$ profiles around the center of the k line. \fref{fig5} 
shows an example, corresponding to a line of sight with $\mu=0.417$, where the 
Stokes $U$ antisymmetric signal is very significant. This antisymmetric signal 
cannot be due to cross-talk between the Stokes parameters, not only because 
the instrumental polarization of CLASP2 is negligible \citep{Song2022SoPh}, 
but also because the observed circular polarization is at the noise level. 
Based on forward modeling calculations, we have found that neither the magnetic 
field nor the $J^\dagger$ symmetry breaking contributions seem to be able to 
produce such antisymmetric shapes in the Stokes $U$ profile of the \ion{Mg}{2} 
k line. The only way we have found to fit these profiles (under our assumption
of \gls*{1d} plane-parallel model atmosphere) is via a stratification in the 
horizontal component of the bulk velocity.

In our inversion of such Stokes profiles we fixed the longitudinal component 
of the macroscopic velocity from the inversion of Stokes $I$. We then performed 
another inversion cycle to get the magnetic field vector and $J^\dagger$, as well 
as the velocity component perpendicular to the LOS ($v_\perp$) and its 
azimuth in the plane perpendicular to the LOS ($\phi_{v_\perp}$). We 
inferred $B_\parallel$ with three nodes, $v_\perp$ and $\phi_{v_\perp}$ with 
two nodes, and $B_\perp$, $\phi_{B_\perp}$, $r_1$, $i_1$, $r_2$, and $i_2$ with 
one node. Although $v_\perp$ and $\phi_{v_\perp}$ have two nodes, one of them 
is located in the lower atmosphere and was fixed to zero, so only the node in 
the upper chromosphere is free to change in the inversion. We opted for this 
strategy because the impact of these components on the polarization comes from 
their spatial gradients rather than from their absolute values.

As shown by the solid curves of \fref{fig5}, the above-mentioned inversion gives 
a very good fit to the observed Stokes $U$ profile. In the top-right panel of 
the figure the blue curve shows the component of the horizontal velocity 
perpendicular to the plane containing the local vertical and the LOS 
($v_y$). The change in $v_y$ between \TAUE{-7.0} and \TAUE{-3.0} is about 
5~km/s. If we assume that the vertical extension of the chromosphere is about 
1500~km (a typical value in semi-empirical models, such as those by 
\citealt{Fontenla1993ApJ}) then the gradient of $v_y$ with height is about 
3.3~m/s/km. In the same panel of \fref{fig5}, the black solid curve shows the
component of the horizontal velocity in the plane containing the local vertical 
and the LOS ($v_x$). The inferred $v_x$, which can modify the amplitude 
of the Stokes $Q$ troughs, is rather constant in the chromosphere of the model 
resulting from the inversion. The inferred $J^\dagger$ values are of the order 
of one percent, or less. The inferred $B_\perp$ is about 100~G, slightly larger 
than the values expected for the quiet Sun chromosphere. This value is almost 
in the saturation regime for the Hanle effect in the Mg {\sc ii} k line, showing 
a quite large error bar (about $\pm50$~G), indicating that the cost function 
is not very sensitive to changes in $B_\perp$. The inferred longitudinal 
magnetic field is about -20~G at \TAUE{-7.0}, producing circular polarization 
signals barely above the noise level.


\section{Degeneracies and ambiguities}\label{sec5}

\begin{figure*}[htp]
  \center
  \includegraphics[width=0.95\textwidth]{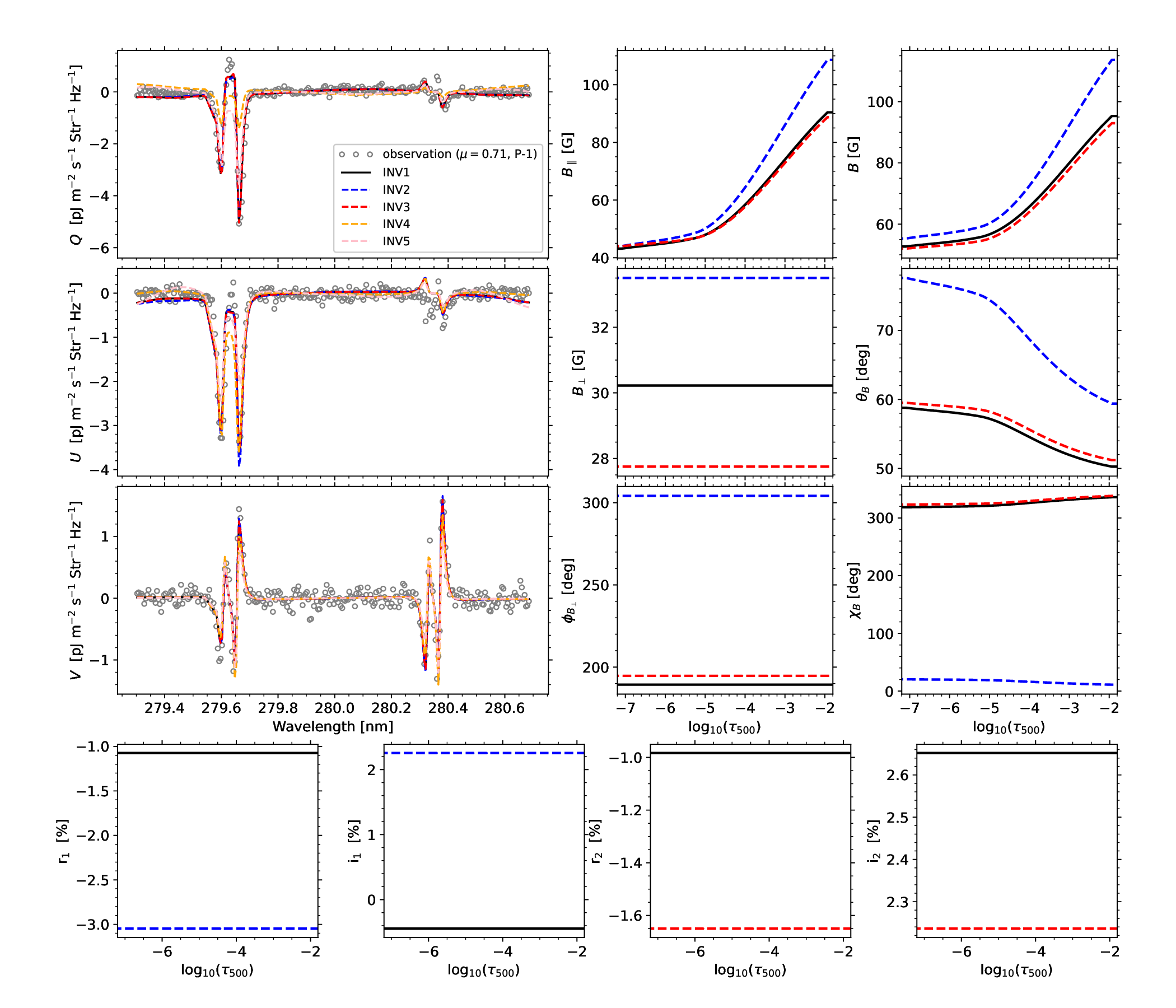}
  \caption{From top to bottom (top three panels in the left column), Stokes 
  $Q$, $U$, and $V$ profiles, respectively. The open circles correspond to the 
  temporally and spatially averaged observed Stokes profiles. From left to right 
  and top to bottom (top three panels in the second and third columns), 
  longitudinal component of the magnetic field, magnetic field strength, 
  transversal component of the magnetic field, magnetic field inclination with 
  respect to the local vertical, magnetic field azimuth in the plane perpendicular 
  to the LOS, and magnetic field azimuth in the local vertical reference 
  frame, respectively. From left to right (bottom row), the real part of the 
  $Q=1$ tensor, the imaginary part of the $Q=1$ tensor, the real part of the 
  $Q=2$ tensor, and the imaginary part of the $Q=2$ tensor characterizing the 
  lack of axial symmetry, respectively. The solid black, and dashed blue, red, 
  orange and pink curves correspond to the fit and inferred quantities from the 
  inversion when including different combinations of horizontal inhomogeneity 
  parameters as free parameters (see legend in the top left panel and see text for the
  description of the labels). The inferred 
  quantities for the orange and pink cases are not shown because they do not fit 
  the observations. This case corresponds to a region at the edge of the plage 
  ($\mu=0.712$ in \fref{fig1}).}
  \label{fig6}
\end{figure*}

\begin{figure*}[htp]
  \center
  \includegraphics[width=0.95\textwidth]{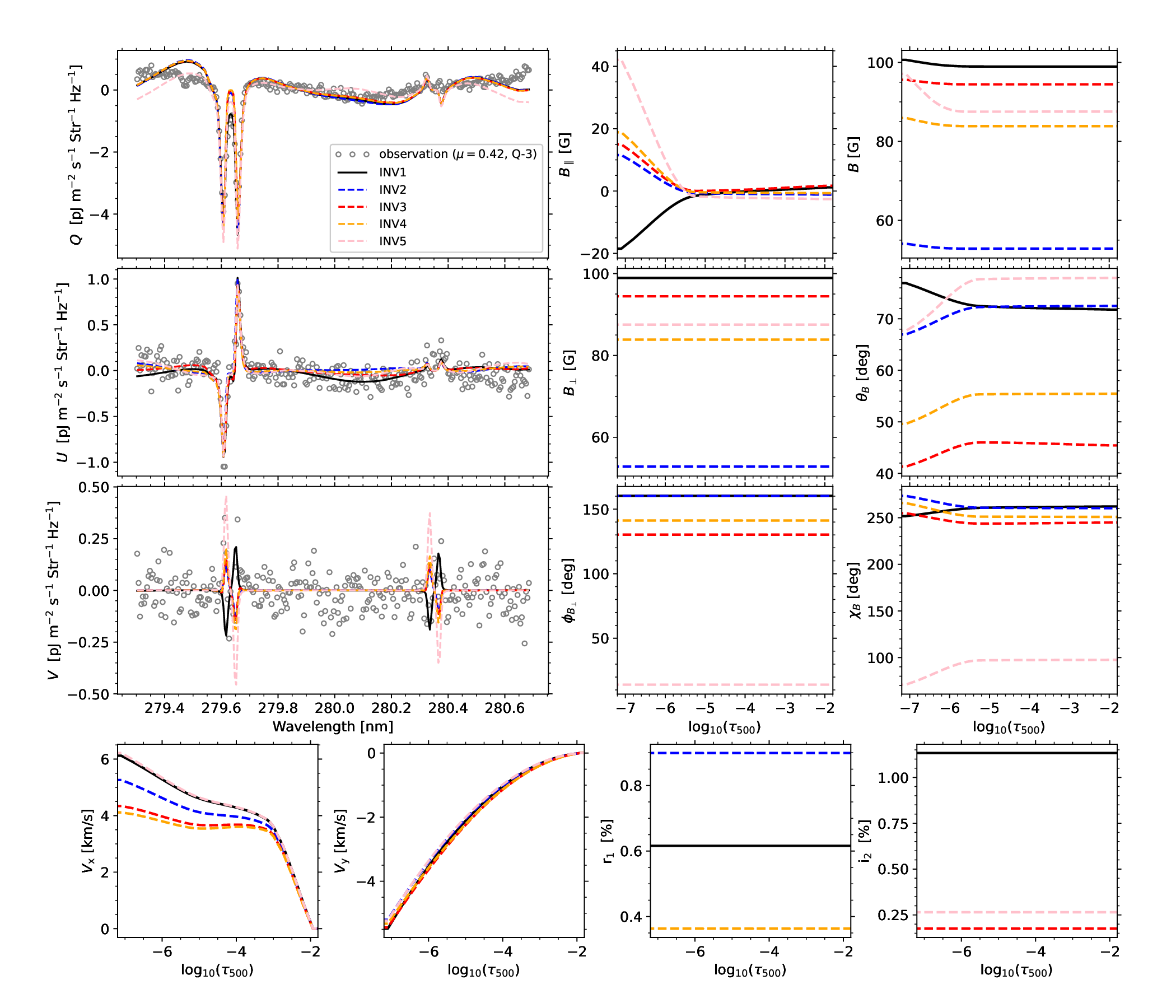}
  \caption{From top to bottom (top three panels in the left column), Stokes 
  $Q$, $U$, and $V$ profiles, respectively. The open circles correspond to 
  the temporally and spatially averaged observed Stokes profiles. From left to 
  right and top to bottom (top three panels in the second and third columns), 
  longitudinal component of the magnetic field, magnetic field strength, 
  transversal component of the magnetic field, magnetic field inclination with 
  respect to the local vertical, magnetic field azimuth in the plane perpendicular 
  to the LOS, and magnetic field azimuth in the local vertical reference 
  frame, respectively. From left to right (bottom row), horizontal component of 
  the plasma velocity in the plane containing the LOS and the local 
  vertical, horizontal component of the plasma velocity perpendicular to the 
  same plane and the vertical, the real part of the $Q=1$ tensor, and the 
  imaginary part of the $Q=2$ tensor characterizing the lack of axial symmetry, 
  respectively. The solid black, and dashed blue, red, orange and pink curves 
  correspond to the fit and inferred quantities from the inversion when 
  including different combinations of horizontal inhomogeneity parameters in 
  the inversion (see legend in the top left panel and see text for the description
  of the labels). This case corresponds to a 
  region in the quiet Sun ($\mu=0.417$ in \fref{fig1}).}
  \label{fig7}
\end{figure*}

The $J^\dagger$ parameters in Eqs. (1) have a physical meaning, namely the 
symmetry breaking contribution from the presence of horizontal inhomogeneities 
in the solar atmosphere and macroscopic velocity gradients. The impact of the 
non-magnetic causes of symmetry breaking can be investigated through spectral 
synthesis calculations with codes like PORTA \citep{Stepan2013A&A} that
accounts for the effects of horizontal radiative transfer 
\citep[e.g.,][]{Jaumeetal2021}. However, \gls*{tic} is a \gls*{1d} code, in 
which $J^\dagger$ was introduced as an ad-hoc parameter aimed at mimicking the 
missing physics. It is thus natural to ask if degeneracy and trade-off exist 
between the $J^\dagger$ contributions, both among themselves and with the 
magnetic field vector. The quick answer to this question is yes. The radiation 
field is calculated in the reference frame with the quantization axis along the 
local vertical and then transformed to the reference frame with the quantization 
axis along the magnetic field \citep{LL04}. This transformation, which consists 
of a linear rotation with Euler angles, makes it possible that different sets
of $J^\dagger$ transform into the same set of rotated tensors, especially given 
that the magnetic field direction is changing during the inversion steps. 
Moreover, it can also happen that some of the components of $J^\dagger$ impact 
the Stokes profiles in a similar (or opposite) way as some of the components 
of the magnetic field. Consequently, it is of critical importance to investigate 
how the degeneracy affects the inferred magnetic field vector. To this end, we 
have made a number of numerical experiments by applying HanleRT-TIC to some of 
CLASP2 Stokes profiles, considering different subsets of $J^\dagger$.

Figure~\ref{fig6} shows the result of several inversions of P-1
in \fref{fig2} for different combinations of $J^\dagger$. The black 
solid curves show the inversion results obtained using the full $J^\dagger$ 
contributions (hereafter INV1), i.e., including both ${J^\dagger}^2_1$ and 
${J^\dagger}^2_2$ with $r_1$, $i_1$, $r_2$, and $i_2$. The blue dashed curves 
show the inversion results including only ${J^\dagger}^2_1$ (hereafter INV2) 
with $r_1$ and $i_1$. The red dashed curves show the inversion results including 
only ${J^\dagger}^2_2$ (hereafter INV3) with $r_2$ and $i_2$. The orange dashed 
curves show the inversion results including only the real components of 
$J^\dagger$ (hereafter INV4) with $r_1$ and $r_2$. Finally, the pink
dashed curves show the inversion results including only the imaginary 
components of $J^\dagger$ (hereafter INV5) with $i_1$ and $i_2$.

INV1, INV2, and INV3 provide good fits to the observations, while those of 
INV4 and INV5 clearly do not (and thus the inferred model atmospheres are not 
shown in \fref{fig6}). The three good fits return rather similar $B_\parallel$, 
$B_\perp$, and $B$, within a few gauss. However, the inferred $\phi_{B_\perp}$ 
shows significant differences. INV1 and INV3 return $\sim190^\circ$, while INV3 
returns $\sim305^\circ$. If we look at the directions in the local reference 
frame, we find that the inclinations $\theta_B$ are relatively similar, within 
10--20 degrees. A significant difference is found in the azimuth $\chi_B$, for 
which INV1 and INV3 return about $320^\circ$ (or $-40^\circ$) and INV2 returns 
about $20^\circ$. This discrepancy is likely due to intrinsic ambiguities of 
the Hanle effect. In contrast to the $180^\circ$ ambiguity characteristic of the
Zeeman effect in the LOS reference frame, the ambiguity of the Hanle 
effect is related to the magnetic field vector in the local vertical frame. 
Analyzing this ambiguity for the \ion{Mg}{2} k line is more complicated due to 
the strong impact of the \gls*{mo} and \gls*{prd} effects on the linear 
polarization in the wings. In essence, the ambiguity of the Hanle effect 
consists on different combinations of $\theta_B$ and $\chi_B$ resulting
in the same $Q$ and $U$. Even with these ambiguities, the inversions still 
confirms a $B_\parallel$ of about 45$\pm$5~G in the middle and upper 
chromosphere and a $B_\perp$ of about 30$\pm$4~G. Both $\phi_{B_\perp}$ and 
$\chi_B$ are affected by ambiguities. We also note that the inferred magnetic 
field vector is very similar between INV1 and INV3. The difference in the 
inferred values is much more important for $J^\dagger$, showing the degeneracy 
between the different components. INV3 returns larger values to compensate for 
the lack of ${J^\dagger}^2_1$, while INV2 returns larger values to compensate 
for the lack of ${J^\dagger}^2_2$.

Figure~\ref{fig7} shows the result of the same test for Q-2 in 
\fref{fig5}. INV1, INV3, INV4, and INV5 return a $B_\perp$ around 90~G, but INV2 
returns about 50~G instead. The inferred $B_\parallel$ shows some differences, 
especially in the uppermost layers of the model, with INV1 returning about 
-20~G in the upper chromosphere, INV2, INV3, and INV4 about 20~G, and INV5 
about 40~G. However, among all the inversions only INV1 can fit the Stokes 
$Q$ signal of the \ion{Mg}{2} k line at its center. The other inversions 
could either get stuck in a local minimum or they could lack the degrees of
freedom in $J^\dagger$ necessary to correctly fit the observation. In this 
case we also find the ambiguity in $\phi_{B_\perp}$ and $\chi_B$. All 
inversions are able to successfully fit the antisymmetric Stokes $U$ profile 
in the k line, returning an almost identical stratification of $v_y$ and 
similar stratifications of $v_x$. All of them retrieve an almost constant 
$v_x$ in the chromosphere and about a change of 5~km/s between lower and 
upper chromosphere in $v_y$. This is a confirmation of the $v_y$ stratification 
being the responsible of the antisymmetric shape of the Stokes $U$ profile 
in the \ion{Mg}{2} k line.

We have performed the same test for Q-3 and Q-4 in \frefs{fig3}--\ref{fig4}, respectively, 
and the results can be found in the Appendix, in \frefs{figb1}--\ref{figb2}. 
Our tests confirm the degeneracy between the ${J^\dagger}^2_1$ and 
${J^\dagger}^2_2$. The inferred $B_\parallel$, $B_\perp$, and $B$ are similar 
within a few gauss, while $\phi_{B_\perp}$ and $\chi_B$ show differences 
likely due to the Hanle-effect ambiguities.


\section{Summary and conclusions} \label{sec6}

While the intensity of the \ion{Mg}{2} h \& k lines allows us to infer 
information on the thermodynamic and dynamic properties of the solar 
chromosphere, their polarization signals encode information on the magnetic 
field all the way up from the upper photosphere to almost the base of the 
corona. The polarization in these strong resonance lines result from the 
joint action of scattering processes and the Hanle, Zeeman, and \gls*{mo} 
effects caused by the presence of magnetic fields. In particular, a rigorous 
modeling of their linear polarization signals requires solving the non-LTE 3D 
radiative transfer problem accounting for the radiatively induced atomic 
level polarization, \gls*{prd} effects, and quantum interference between 
the magnetic sublevels pertaining to their upper $J$-levels, a very challenging 
unsolved problem, although some progress toward this goal has been recently
made \citep{Benedusi2023JCoPh}. The fact that the solar atmosphere is 
horizontally inhomogeneous and dynamic implies that the horizontal radiative 
transfer can break the axial symmetry of the radiation field that pumps the 
atoms at each spatial point within the atmospheric plasma, without the need 
of any inclined magnetic field. Therefore, in general, we have magnetic and 
non-magnetic causes of axial symmetry breaking.

Our \gls*{tic} solves the radiative transfer problem taking into account 
all the above-mentioned mechanisms (i.e., \gls*{prd}, $J$-state interference, 
Hanle, Zeeman and \gls*{mo} effects), but assuming 1D plane-parallel geometry 
(i.e., ignoring the effects of horizontal radiative transfer). In order to 
take into account in our Stokes inversions the possibility of non-magnetic 
causes of axial symmetry breaking (e.g., because of the presence of horizontal 
inhomogeneities in the temperature and density of the plasma), we have introduced 
in \gls*{tic} ad-hoc components in the radiation field tensor that quantifies 
the axial symmetry breaking of the pumping radiation field. In this paper we
have shown that such $J^\dagger$ contributions, which are additional Stokes 
inversion parameters, allow us to successfully fit a variety of Stokes profiles 
from the CLASP2 observations. For demonstrative purposes, we selected four 
representative types of the variety of CLASP2 Stokes profiles, including a 
location in the plage target (P-1, where Stokes-$V$ is significant but Stokes 
$Q$ and $U$ are negligible in the far wings) and three locations (Q-1, Q-2, and Q-3)
in the quiet Sun target. At Q-1, Stokes $Q$ is 
significant in the far wings but Stokes $U$ is negligible, at Q-2 
both Stokes $Q$ and $U$ are above the noise level, and at Q-3 we find
antisymmetric Stokes $U$ profiles around the center of the k line.  
   
We have found that the inclusion of $J^\dagger$ is necessary for fitting some 
of the CLASP2 Stokes profiles (P-1 and Q-2). 
In addition, we have identified a clear degeneracy between ${J^\dagger}^2_1$ 
and ${J^\dagger}^2_2$. It is important to emphasize that these ad-hoc 
contributions to the radiation field tensors are accounted for in a 
\gls*{3d} forward solver \citep{Stepan2013A&A,Benedusi2023JCoPh} and for some
lines can be estimated from the continuum illumination \citep[][]{Zeuner2020ApJ}, 
although this method would not be suitable for the \ion{Mg}{2} h \& k lines. 
In order to address this challenge in an inversion a numerical code is in 
development following the guidelines in \cite{Stepan2022A&A}, but the 
methodology is restricted, in principle, to spectral lines in the complete 
frequency redistribution regime (i.e., without \gls*{prd} effects).

Interestingly, we have found that a gradient in the component of the horizontal 
velocity perpendicular to the plane containing the local vertical and the 
LOS ($v_y$) is needed to produce and fit one of our selected sets of 
Stokes profiles, namely the one showing an antisymmetric Stokes $U$ profile 
around the \ion{Mg}{2} k line center. A difference of about 5~km/s between 
\TAUE{-7.0} and -3.0 is found in the inversion of this Stokes profile, shown 
in \fref{fig5}. It is noteworthy that this kind of profile is only observed at 
a few positions along the CLASP2 slit. This suggests that in the quiet Sun 
chromosphere the gradient of $v_y$ is, generally, smaller than this value.

By performing several inversions with different subsets of $J^\dagger$ as free 
parameters, we have studied the degeneracy we have identified among the 
$J^\dagger$ components themselves, and with the magnetic field vector. For 
the four analyzed types of Stokes profiles, $B_\parallel$, $B_\perp$, and $B$ 
are found within a few gauss, typically $\pm10$~G for all inversions (when they 
are able to fit the observations at all). The differences in $\theta_B$,
$\phi_{B_\perp}$, and $\chi_B$ are more significant and they could be due to 
the ambiguities of the Hanle effect. Moreover, the lack of clear circular 
polarization signals, as in Q-1 and Q-2, favors 
a wider range of compatible solutions. The analysis of P-1 
shown in \fref{fig6} is indicative of how the access to circular polarization 
signals allows a much better constrain of $B_\parallel$ and, consequently, of the 
magnetic field vector, except for the ambiguities in the azimuths.

We have demonstrated that \gls*{tic} is able to infer the vector magnetic field 
of the solar chromosphere from the Stokes profiles of the \ion{Mg}{2} h \& k 
lines, including the complex physical ingredients that are needed for their 
modeling. For example, in the quiet region pixels, where no circular 
polarization signal is detected, the magnetic field strength in the
upper chromosphere varies between 1 and 20 gauss.  

Of course, the accuracy of the retrieved magnetic field vector is limited by 
the polarimetric accuracy of the observations. Moreover, the heavy computational 
requirement of these non-LTE Stokes inversions severely limits its applicability 
to a large dataset. A faster inversion may be possible by applying clustering 
methods to select representative profiles \citep{SainzDalda2019ApJ}, by 
training convolutional neural networks to speed-up the computation of response 
functions \citep{Centeno2022ApJ}, or by extending other machine
learning techniques \citep{VicenteArevalo2022ApJ,AsensioRamos2023LRSP} 
to the more general non-LTE problem with atomic polarization.

Finally, we find it important to emphasize again  that the results of the 
previous works on the analysis of the CLASP2 data 
\citep{Ishikawa2021SA,Li2023ApJ,AfonsoDelgado2023ApJ} and those of the present 
investigation highlight the potential of a new space mission with CLASP2-like 
capabilities for the study of the magnetic field in the solar chromosphere.


\acknowledgements

We gratefully acknowledge the financial support from the European Research 
Council (ERC) under the European Union's Horizon 2020 research and innovation 
programme (Advanced Grant agreement No. 742265). T.P.A.'s participation in the 
publication is part of the Project RYC2021-034006-I, funded by 
MICIN/AEI/10.13039/501100011033, and the European Union “NextGenerationEU”/RTRP. 
T.P.A. and J.T.B. acknowledge support from the Agencia Estatal
de Investigación del Ministerio de Ciencia, Innovación y Universidades 
(MCIU/AEI) under grant ``Polarimetric Inference of Magnetic Fields'' and the 
European Regional Development Fund (ERDF) with reference
PID2022-136563NB-I00/10.13039/501100011033.
H.L. acknowledges the support from the National Key R\&D Program of China 
(2021YFA1600500, 2021YFA1600503) and the National Natural Science
Foundation of China under grant No. 12473051.
CLASP2 is an international partnership between NASA/MSFC, NAOJ, JAXA, IAC, 
and IAS; additional partners include ASCR, IRSOL, LMSAL, and the University of 
Oslo.


\appendix

\counterwithin{figure}{section}


\section{Spatial average of the Stokes profiles}\label{app-avg}

\begin{figure}
  \center
  \includegraphics[width=0.6\textwidth]{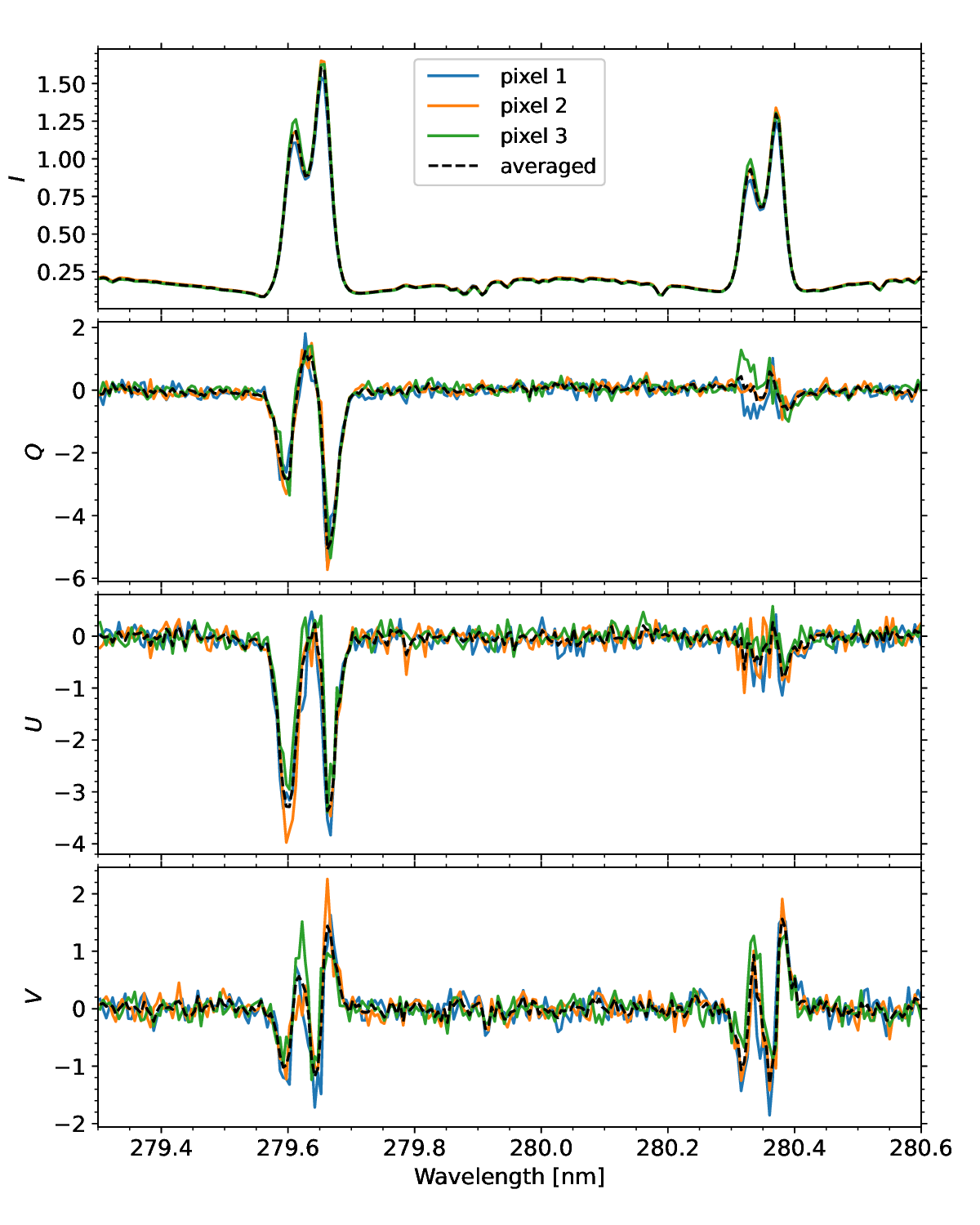}
  \caption{From top to bottom, the intensity $I$ and Stokes $Q$, $U$, and $V$ profiles
  for the spatial average at $\mu = 0.712$ (dashed black curve; see \fref{fig2}). The colored
  solid curves show the profiles for the individual pixels included in the average.
  The y axis units are the same as those in \fref{fig2}.}
  \label{figa1}
\end{figure}

Figure~\ref{figa1} shows the spatially averaged profile at $\mu=0.712$
(P-1, see section~\ref{sec4}), as well as the Stokes profiles for the three
individual pixels of the observation included in the average.

\section{Degeneracy tests}

\begin{figure*}[htp]
  \center
  \includegraphics[width=0.95\textwidth]{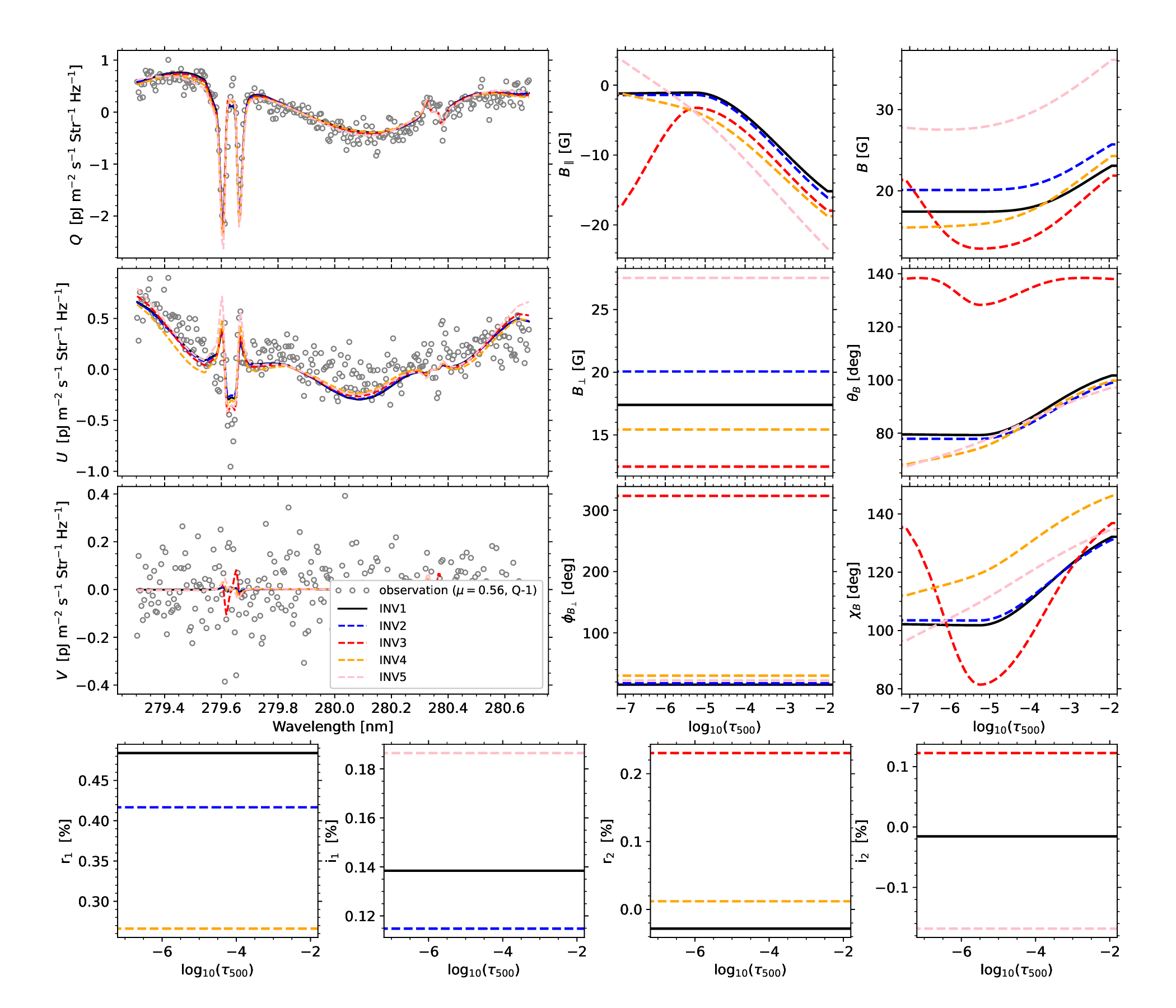}
  \caption{Same as \fref{fig6}, but for a region in the quiet Sun with 
  $\mu=0.563$ (see \fref{fig1}).}
  \label{figb1}
\end{figure*}

\begin{figure*}[htp]
  \center
  \includegraphics[width=0.95\textwidth]{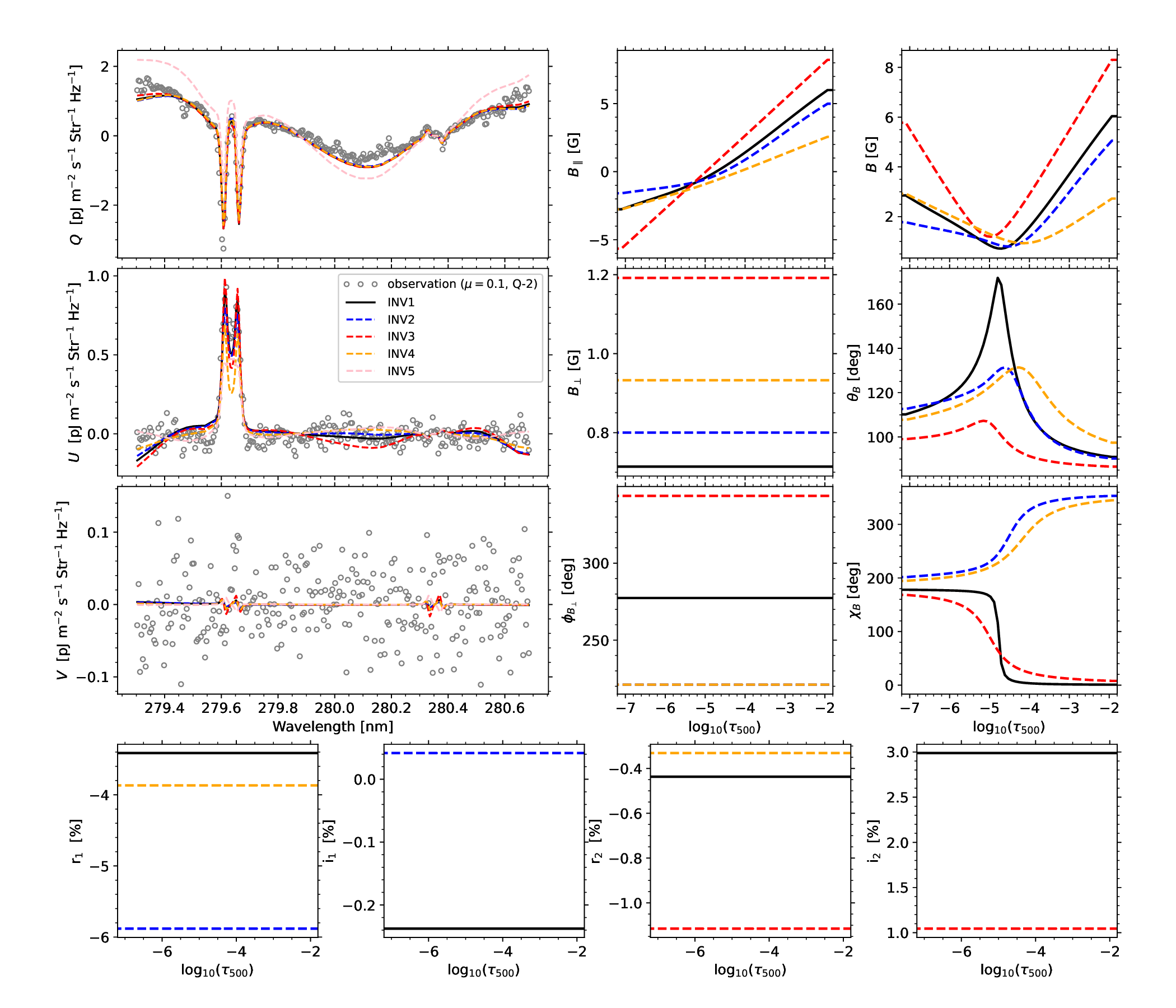}
  \caption{Same as \fref{fig6}, but for a region in the quiet Sun with 
  $\mu=0.101$ (see \fref{fig1}).}
  \label{figb2}
\end{figure*}

In \frefs{figb1} and \ref{figb2} we show the degeneracy analysis 
(see Sect.~\ref{sec5}) of the profiles in \frefs{fig3} and \ref{fig4}, 
respectively. As with the results shown in Sect.~\ref{sec5}, the inferred 
$B_\parallel$, $B_\perp$, and $B$ are usually within a few gauss. However, 
noticeable differentces are found for $\phi_{B_\perp}$, $\theta_B$, and 
$\chi_B$, possibly due to the ambiguities in the Hanle effect.

\bibliography{clasp2limb}{}

\begin{thebibliography}{}
\expandafter\ifx\csname natexlab\endcsname\relax\def\natexlab#1{#1}\fi
\providecommand{\url}[1]{\href{#1}{#1}}
\providecommand{\dodoi}[1]{doi:~\href{http://doi.org/#1}{\nolinkurl{#1}}}
\providecommand{\doeprint}[1]{\href{http://ascl.net/#1}{\nolinkurl{http://ascl.net/#1}}}
\providecommand{\doarXiv}[1]{\href{https://arxiv.org/abs/#1}{\nolinkurl{https://arxiv.org/abs/#1}}}

\bibitem[{{Afonso Delgado} {et~al.}(2023){Afonso Delgado}, {del Pino
  Alem{\'a}n}, \& {Trujillo Bueno}}]{AfonsoDelgado2023ApJ}
{Afonso Delgado}, D., {del Pino Alem{\'a}n}, T., \& {Trujillo Bueno}, J. 2023,
  \apj, 954, 218, \dodoi{10.3847/1538-4357/ace4c8}

\bibitem[{{Alsina Ballester} {et~al.}(2016){Alsina Ballester}, {Belluzzi}, \&
  {Trujillo Bueno}}]{AlsinaBallester2016ApJ}
{Alsina Ballester}, E., {Belluzzi}, L., \& {Trujillo Bueno}, J. 2016, \apjl,
  831, L15, \dodoi{10.3847/2041-8205/831/2/L15}

\bibitem[{{Asensio Ramos} {et~al.}(2023){Asensio Ramos}, {Cheung}, {Chifu}, \&
  {Gafeira}}]{AsensioRamos2023LRSP}
{Asensio Ramos}, A., {Cheung}, M. C.~M., {Chifu}, I., \& {Gafeira}, R. 2023,
  Living Reviews in Solar Physics, 20, 4, \dodoi{10.1007/s41116-023-00038-x}

\bibitem[{{Asensio Ramos} {et~al.}(2008){Asensio Ramos}, {Trujillo Bueno}, \&
  {Land i Degl'Innocenti}}]{AsensioRamos2008ApJ}
{Asensio Ramos}, A., {Trujillo Bueno}, J., \& {Land i Degl'Innocenti}, E. 2008,
  \apj, 683, 542, \dodoi{10.1086/589433}

\bibitem[{{Belluzzi} \& {Trujillo Bueno}(2012)}]{Belluzzi2012ApJa}
{Belluzzi}, L., \& {Trujillo Bueno}, J. 2012, \apjl, 750, L11,
  \dodoi{10.1088/2041-8205/750/1/L11}

\bibitem[{{Benedusi} {et~al.}(2023){Benedusi}, {Riva}, {Zulian},
  {{\v{S}}t{\v{e}}p{\'a}n}, {Belluzzi}, \& {Krause}}]{Benedusi2023JCoPh}
{Benedusi}, P., {Riva}, S., {Zulian}, P., {et~al.} 2023, Journal of
  Computational Physics, 479, 112013, \dodoi{10.1016/j.jcp.2023.112013}

\bibitem[{{Centeno} {et~al.}(2022){Centeno}, {Flyer}, {Mukherjee}, {Egeland},
  {Casini}, {del Pino Alem{\'a}n}, \& {Rempel}}]{Centeno2022ApJ}
{Centeno}, R., {Flyer}, N., {Mukherjee}, L., {et~al.} 2022, \apj, 925, 176,
  \dodoi{10.3847/1538-4357/ac402f}

\bibitem[{{de la Cruz Rodr{\'\i}guez} \& {van Noort}(2017)}]{delaCruz2017SSRv}
{de la Cruz Rodr{\'\i}guez}, J., \& {van Noort}, M. 2017, \ssr, 210, 109,
  \dodoi{10.1007/s11214-016-0294-8}

\bibitem[{{del Pino Alem{\'a}n} {et~al.}(2016){del Pino Alem{\'a}n}, {Casini},
  \& {Manso Sainz}}]{Tanausu2016ApJ}
{del Pino Alem{\'a}n}, T., {Casini}, R., \& {Manso Sainz}, R. 2016, \apjl, 830,
  L24, \dodoi{10.3847/2041-8205/830/2/L24}

\bibitem[{{del Pino Alem{\'a}n} {et~al.}(2020){del Pino Alem{\'a}n}, {Trujillo
  Bueno}, {Casini}, \& {Manso Sainz}}]{Tanausu2020ApJ}
{del Pino Alem{\'a}n}, T., {Trujillo Bueno}, J., {Casini}, R., \& {Manso
  Sainz}, R. 2020, \apj, 891, 91, \dodoi{10.3847/1538-4357/ab6bc9}

\bibitem[{{del Pino Alem{\'a}n} {et~al.}(2018){del Pino Alem{\'a}n}, {Trujillo
  Bueno}, {{\v S}t{\v e}p{\'a}n}, \& {Shchukina}}]{delPinoetal2018}
{del Pino Alem{\'a}n}, T., {Trujillo Bueno}, J., {{\v S}t{\v e}p{\'a}n}, J., \&
  {Shchukina}, N. 2018, \apj, 863, 164, \dodoi{10.3847/1538-4357/aaceab}

\bibitem[{{del Toro Iniesta} \& {Ruiz Cobo}(2016)}]{Iniesta2016LRSP}
{del Toro Iniesta}, J.~C., \& {Ruiz Cobo}, B. 2016, Living Reviews in Solar
  Physics, 13, 4, \dodoi{10.1007/s41116-016-0005-2}

\bibitem[{{Fontenla} {et~al.}(1993){Fontenla}, {Avrett}, \&
  {Loeser}}]{Fontenla1993ApJ}
{Fontenla}, J.~M., {Avrett}, E.~H., \& {Loeser}, R. 1993, \apj, 406, 319,
  \dodoi{10.1086/172443}

\bibitem[{{Ishikawa} {et~al.}(2021){Ishikawa}, {Trujillo Bueno}, {del Pino
  Alem{\'a}n}, {Okamoto}, {McKenzie}, {Auch{\`e}re}, {Kano}, {Song}, {Yoshida},
  {Rachmeler}, {Kobayashi}, {Hara}, {Kubo}, {Narukage}, {Sakao}, {Shimizu},
  {Suematsu}, {Bethge}, {De Pontieu}, {Dalda}, {Vigil}, {Winebarger},
  {Ballester}, {Belluzzi}, {{\v{S}}t{\v{e}}p{\'a}n}, {Ramos}, {Carlsson}, \&
  {Leenaarts}}]{Ishikawa2021SA}
{Ishikawa}, R., {Trujillo Bueno}, J., {del Pino Alem{\'a}n}, T., {et~al.} 2021,
  Science Advances, 7, eabe8406, \dodoi{10.1126/sciadv.abe8406}

\bibitem[{{Ishikawa} {et~al.}(2023){Ishikawa}, {Trujillo Bueno}, {Alsina
  Ballester}, {Belluzzi}, {del Pino Alem{\'a}n}, {McKenzie}, {Auch{\`e}re},
  {Kobayashi}, {Okamoto}, {Rachmeler}, \& {Song}}]{Ishikawa2023ApJ}
{Ishikawa}, R., {Trujillo Bueno}, J., {Alsina Ballester}, E., {et~al.} 2023,
  \apj, 945, 125, \dodoi{10.3847/1538-4357/acb64e}

\bibitem[{{Jaume Bestard} {et~al.}(2021){Jaume Bestard}, {Trujillo Bueno},
  {{\v{S}}t{\v{e}}p{\'a}n}, \& {del Pino Alem{\'a}n}}]{Jaumeetal2021}
{Jaume Bestard}, J., {Trujillo Bueno}, J., {{\v{S}}t{\v{e}}p{\'a}n}, J., \&
  {del Pino Alem{\'a}n}, T. 2021, \apj, 909, 183,
  \dodoi{10.3847/1538-4357/abd94a}

\bibitem[{{Lagg} {et~al.}(2017){Lagg}, {Lites}, {Harvey}, {Gosain}, \&
  {Centeno}}]{Lagg2017SSRv}
{Lagg}, A., {Lites}, B., {Harvey}, J., {Gosain}, S., \& {Centeno}, R. 2017,
  \ssr, 210, 37, \dodoi{10.1007/s11214-015-0219-y}

\bibitem[{{Landi Degl'Innocenti} \& {Landolfi}(2004)}]{LL04}
{Landi Degl'Innocenti}, E., \& {Landolfi}, M. 2004, {Polarization in Spectral
  Lines}, Vol. 307 ({Dordrecht:KluwerAcademic}),
  \dodoi{10.1007/978-1-4020-2415-3}

\bibitem[{{Li} {et~al.}(2022){Li}, {del Pino Alem{\'a}n}, {Trujillo Bueno}, \&
  {Casini}}]{Li2022ApJ}
{Li}, H., {del Pino Alem{\'a}n}, T., {Trujillo Bueno}, J., \& {Casini}, R.
  2022, \apj, 933, 145, \dodoi{10.3847/1538-4357/ac745c}

\bibitem[{{Li} {et~al.}(2023){Li}, {del Pino Alem{\'a}n}, {Trujillo Bueno},
  {Ishikawa}, {Alsina Ballester}, {McKenzie}, {Auch{\`e}re}, {Kobayashi},
  {Okamoto}, {Rachmeler}, \& {Song}}]{Li2023ApJ}
{Li}, H., {del Pino Alem{\'a}n}, T., {Trujillo Bueno}, J., {et~al.} 2023, \apj,
  945, 144, \dodoi{10.3847/1538-4357/acb76e}

\bibitem[{{Manso Sainz} \& {Trujillo Bueno}(2011)}]{MansoTrujillo2011}
{Manso Sainz}, R., \& {Trujillo Bueno}, J. 2011, \apj, 743, 12,
  \dodoi{10.1088/0004-637X/743/1/12}

\bibitem[{{Narukage} {et~al.}(2016){Narukage}, {McKenzie}, {Ishikawa},
  {Trujillo-Bueno}, {De Pontieu}, {Kubo}, {Ishikawa}, {Kano}, {Suematsu},
  {Yoshida}, {Rachmeler}, {Kobayashi}, {Cirtain}, {Winebarger}, {Asensio
  Ramos}, {del Pino Aleman}, {{\v{S}}t{\k{e}}p{\'a}n}, {Belluzzi},
  {Larruquert}, {Auch{\`e}re}, {Leenaarts}, \& {Carlsson}}]{Narukage2016SPIE}
{Narukage}, N., {McKenzie}, D.~E., {Ishikawa}, R., {et~al.} 2016, in Society of
  Photo-Optical Instrumentation Engineers (SPIE) Conference Series, Vol. 9905,
  Space Telescopes and Instrumentation 2016: Ultraviolet to Gamma Ray, ed.
  J.-W.~A. {den Herder}, T.~{Takahashi}, \& M.~{Bautz}, 990508,
  \dodoi{10.1117/12.2232245}

\bibitem[{{Rachmeler} {et~al.}(2022){Rachmeler}, {Trujillo Bueno}, {McKenzie},
  {Ishikawa}, {Auch{\`e}re}, {Kobayashi}, {Kano}, {Okamoto}, {Bethge}, {Song},
  {Alsina Ballester}, {Belluzzi}, {del Pino Alem{\'a}n}, {Asensio Ramos},
  {Yoshida}, {Shimizu}, {Winebarger}, {Kobelski}, {Vigil}, {De Pontieu},
  {Narukage}, {Kubo}, {Sakao}, {Hara}, {Suematsu}, {{\v{S}}t{\v{e}}p{\'a}n},
  {Carlsson}, \& {Leenaarts}}]{Rachmeler2022ApJ}
{Rachmeler}, L.~A., {Trujillo Bueno}, J., {McKenzie}, D.~E., {et~al.} 2022,
  \apj, 936, 67, \dodoi{10.3847/1538-4357/ac83b8}

\bibitem[{{Sainz Dalda} {et~al.}(2019){Sainz Dalda}, {de la Cruz
  Rodr{\'\i}guez}, {De Pontieu}, \& {Go{\v{s}}i{\'c}}}]{SainzDalda2019ApJ}
{Sainz Dalda}, A., {de la Cruz Rodr{\'\i}guez}, J., {De Pontieu}, B., \&
  {Go{\v{s}}i{\'c}}, M. 2019, \apjl, 875, L18, \dodoi{10.3847/2041-8213/ab15d9}

\bibitem[{{Song} {et~al.}(2018){Song}, {Ishikawa}, {Kano}, {Yoshida},
  {Tsuzuki}, {Uraguchi}, {Shinoda}, {Hara}, {Okamoto}, {Auch{\`e}re},
  {McKenzie}, {Rachmeler}, \& {Trujillo Bueno}}]{Song2018SPIE}
{Song}, D., {Ishikawa}, R., {Kano}, R., {et~al.} 2018, in Society of
  Photo-Optical Instrumentation Engineers (SPIE) Conference Series, Vol. 10699,
  Space Telescopes and Instrumentation 2018: Ultraviolet to Gamma Ray, ed.
  J.-W.~A. {den Herder}, S.~{Nikzad}, \& K.~{Nakazawa}, 106992W,
  \dodoi{10.1117/12.2313056}

\bibitem[{{Song} {et~al.}(2022){Song}, {Ishikawa}, {Kano}, {McKenzie},
  {Trujillo Bueno}, {Auch{\`e}re}, {Rachmeler}, {Okamoto}, {Yoshida},
  {Kobayashi}, {Bethge}, {Hara}, {Shinoda}, {Shimizu}, {Suematsu}, {De
  Pontieu}, {Winebarger}, {Narukage}, {Kubo}, {Sakao}, {Asensio Ramos},
  {Belluzzi}, {{\v{S}}t{\v{e}}p{\'a}n}, {Carlsson}, {del Pino Alem{\'a}n},
  {Alsina Ballester}, {Vigil}, \& {Leenaarts}}]{Song2022SoPh}
{Song}, D., {Ishikawa}, R., {Kano}, R., {et~al.} 2022, \solphys, 297, 135,
  \dodoi{10.1007/s11207-022-02064-8}

\bibitem[{{Trujillo Bueno} \& {del Pino
  Alem{\'a}n}(2022)}]{TrujilloBueno2022ARA&A}
{Trujillo Bueno}, J., \& {del Pino Alem{\'a}n}, T. 2022, \araa, 60, 415,
  \dodoi{10.1146/annurev-astro-041122-031043}

\bibitem[{{Tsuzuki} {et~al.}(2020){Tsuzuki}, {Ishikawa}, {Kano}, {Narukage},
  {Song}, {Yoshida}, {Uraguchi}, {Okamoto}, {McKenzie}, {Kobayashi},
  {Rachmeler}, {Auchere}, \& {Trujillo Bueno}}]{Tsuzuki2020SPIE}
{Tsuzuki}, T., {Ishikawa}, R., {Kano}, R., {et~al.} 2020, in Society of
  Photo-Optical Instrumentation Engineers (SPIE) Conference Series, Vol. 11444,
  Society of Photo-Optical Instrumentation Engineers (SPIE) Conference Series,
  114446W, \dodoi{10.1117/12.2562273}

\bibitem[{{Vicente Ar{\'e}valo} {et~al.}(2022){Vicente Ar{\'e}valo}, {Asensio
  Ramos}, \& {Esteban Pozuelo}}]{VicenteArevalo2022ApJ}
{Vicente Ar{\'e}valo}, A., {Asensio Ramos}, A., \& {Esteban Pozuelo}, S. 2022,
  \apj, 928, 101, \dodoi{10.3847/1538-4357/ac53b3}

\bibitem[{{{\v{S}}t{\v{e}}p{\'a}n} {et~al.}(2022){{\v{S}}t{\v{e}}p{\'a}n}, {del
  Pino Alem{\'a}n}, \& {Trujillo Bueno}}]{Stepan2022A&A}
{{\v{S}}t{\v{e}}p{\'a}n}, J., {del Pino Alem{\'a}n}, T., \& {Trujillo Bueno},
  J. 2022, \aap, 659, A137, \dodoi{10.1051/0004-6361/202142079}

\bibitem[{{{\v{S}}t{\v{e}}p{\'a}n} {et~al.}(2024){{\v{S}}t{\v{e}}p{\'a}n}, {del
  Pino Alem{\'a}n}, \& {Trujillo Bueno}}]{Stepanetal2024}
---. 2024, arXiv e-prints, arXiv:2407.20926, \dodoi{10.48550/arXiv.2407.20926}

\bibitem[{{{\v{S}}t{\v{e}}p{\'a}n} \& {Trujillo Bueno}(2013)}]{Stepan2013A&A}
{{\v{S}}t{\v{e}}p{\'a}n}, J., \& {Trujillo Bueno}, J. 2013, \aap, 557, A143,
  \dodoi{10.1051/0004-6361/201321742}

\bibitem[{{Zeuner} {et~al.}(2024){Zeuner}, {del Pino Alem{\'a}n}, {Trujillo
  Bueno}, \& {Solanki}}]{Zeuneretal2024}
{Zeuner}, F., {del Pino Alem{\'a}n}, T., {Trujillo Bueno}, J., \& {Solanki},
  S.~K. 2024, \apj, 964, 10, \dodoi{10.3847/1538-4357/ad26f9}

\bibitem[{{Zeuner} {et~al.}(2020){Zeuner}, {Manso Sainz}, {Feller}, {van
  Noort}, {Solanki}, {Iglesias}, {Reardon}, \& {Mart{\'\i}nez
  Pillet}}]{Zeuner2020ApJ}
{Zeuner}, F., {Manso Sainz}, R., {Feller}, A., {et~al.} 2020, \apjl, 893, L44,
  \dodoi{10.3847/2041-8213/ab86b8}

\end{thebibliography}
\bibliographystyle{aasjournal}

\end{document}